\documentclass[aps,prd,reprint,superscriptaddress,nofootinbib,floatfix]{revtex4-2}
\usepackage[T1]{fontenc}
\usepackage{amsmath,amssymb,bm,graphicx,booktabs}
\usepackage[colorlinks=true,linkcolor=blue,citecolor=blue,urlcolor=blue]{hyperref}
\hypersetup{pdftitle={Meson-isospectral pair interactions and the residual three-quark force},pdfauthor={Jongheon Baek, Gihwan Nam, Gyucheol Sim}}
\newcommand{\hc}{\hbar c}
\newcommand{\MeV}{\mathrm{MeV}}
\newcommand{\fm}{\mathrm{fm}}

\newcommand{\ket}[1]{|#1\rangle}
\newcommand{\bra}[1]{\langle#1|}
\newcommand{\CC}{\mathcal C}
\graphicspath{{figures/}}
\newcommand{\DirectThreeTrain}{21.19}
\newcommand{\DirectThreeTest}{9.44}

\newcommand{\MassFitAction}{1.360}

\newcommand{\RefinedActionTwo}{0.861}
\newcommand{\RefinedActionThree}{0.753}
\newcommand{\RefinedActionFour}{0.728}
\newcommand{\RefinedActionFive}{0.726}
\newcommand{\RefinedSourceRMS}{201.32}
\newcommand{\RefinedFiveResidualRMS}{146.16}
\newcommand{\RegGapTwo}{26.06}
\newcommand{\RegGapThree}{59.48}
\newcommand{\RegGapFour}{93.81}
\newcommand{\RegMassFourTrain}{0.96}
\newcommand{\RegMassFourTest}{15.89}
\newcommand{\RegGapRefineMax}{0.1455}
\newcommand{\RegGapDifferenceRefineMax}{0.1485}
\newcommand{\RegGapInitialMax}{1.658}
\newcommand{\RefinedNSInitialChange}{0.003844}
\newcommand{\RefinedNSLastChange}{0.001165}
\newcommand{\RefinedOmegaLastChange}{0.000381}
\newcommand{\JointGroundTrain}{1.61}
\newcommand{\JointGroundTest}{10.69}
\newcommand{\JointGapTrain}{1.39}
\newcommand{\JointGapTest}{8.12}
\newcommand{\JointActionAll}{2.372}
\newcommand{\JointActionFloor}{0.043}
\newcommand{\JointActionCal}{2.342}
\newcommand{\JointCalSourceRMS}{292.07}
\newcommand{\JointCalResidualRMS}{684.12}
\newcommand{\JointStrongGroundTest}{8.68}
\newcommand{\JointStrongGapTest}{3.23}
\newcommand{\JointStrongActionAll}{1.959}
\newcommand{\JointXiCCGap}{-0.536}
\newcommand{\JointGapRefineMax}{0.0449}
\newcommand{\JointGroundRefineMax}{0.0380}
\newcommand{\JointGapInitialMax}{1.0906}
\newcommand{\JointActionNSInitialChange}{0.015900}
\newcommand{\JointActionNSLastChange}{0.001387}
\newcommand{\ShortActionAll}{0.711}

\newcommand{\ShortResidualRMS}{143.09}
\newcommand{\ShortNSInitialChange}{0.000445}
\newcommand{\ShortNSLastChange}{0.001262}

\newcommand{\ShortOmegaAtThirteen}{0.646954}
\newcommand{\ShortOmegaAtNineteen}{0.591794}
\newcommand{\ShortOmegaAtTwentyOne}{0.592270}
\begin{document}
\title{Meson-isospectral pair interactions and the residual three-quark force}
\author{Jongheon Baek}
\email{bjh6941@yonsei.ac.kr}
\affiliation{Department of Physics and Institute of Physics and Applied Physics, Yonsei University, Seoul 03722, Korea}
\author{Gihwan Nam}
\affiliation{Department of Physics, Yonsei University, Seoul 03722, South Korea}
\author{Gyucheol Sim}
\affiliation{Department of Physics Education, Seoul National University, Seoul 08826, South Korea}
\date{October 21, 2026}
\begin{abstract}
Meson-isospectral pair rotations can closely reproduce low-lying baryon
energies while producing a substantial connected action under their
consistent three-body embedding. Starting from an independently calibrated
S-wave meson model, we compare ground-mass fits, joint ground-level and gap
fits, and direct minimization of the connected remainder. With eighteen
shared coefficients, a four-packet joint fit reproduces the nine calibration
ground masses and first gaps with RMS errors of $\JointGroundTrain$ and
$\JointGapTrain$ MeV, respectively. The excluded ground masses and gaps
have RMS errors of $10.7$ and $8.1$ MeV in the same enlarged radial space.
Its connected action on the eighteen calibrated inputs is $\JointActionCal$
times the unrotated source action. Eigenvalue agreement and removal of the connected interaction
thus impose different tests. Separately optimizing the action with five
packets gives a normalized residual of $\ShortActionAll$ for an
intermediate-distance fifth packet; a long-distance extension also leaves
a substantial remainder. An eliminated-state construction realizes the
central color--color structure of an intermediate-configuration mechanism
together with correlated pair dressing. These results favor a connected
operator basis with three fitted strengths as an economical description
within the tested pair family,
and quantify its relation to the representation freedom of equivalent
few-body Hamiltonians.
\end{abstract}
\maketitle

\section{Introduction}
Meson and baryon spectroscopy constrain a constituent interaction in different
color and many-particle environments. An explicit connected three-quark term
can improve a meson-calibrated baryon spectrum, as in the operator construction
of Ref.~\cite{Baek2026ThreeBody}. Short-range quark three-body forces have
also been applied to the flavor dependence of baryon
spectra~\cite{Baek2026FlavorDependency}. The inverse question is how much of
this spectral improvement can be reproduced by changing the nonlocal structure of
the pair interaction while maintaining the meson constraints. A numerical
answer requires specifying the admissible pair transformations, their color
extension, and the subtraction defining the connected remainder.

The use of one constituent interaction across mesons and baryons has a long
history. A common light-hadron parameter set was obtained by modifying the
baryon-sector instanton interaction~\cite{BrauSemaySilvestreBrac2002}, while
Faddeev calculations tested a constituent interaction already applied to
meson spectra and baryon--baryon dynamics~\cite{ValcarceGarcilazoVijande2005}.
Quark many-body terms also arise in several distinct dynamical settings,
including instanton-induced interactions in strange
systems~\cite{OkaTakeuchi1991} and color-dependent confinement
models~\cite{PepinStancu2002,PappStancu2003}. In the static limit, lattice
calculations resolve a three-quark confining geometry~\cite{TakahashiEtAl2002},
and perturbative QCD contains a genuine three-body contribution at
next-to-next-to-leading order~\cite{BrambillaGhiglieriVairo2010}.
These examples motivate testing the information carried by a connected term
once the meson interaction and its embedding have been fixed. Here its
finite-range constituent form serves as a reference for that test.

The underlying representation freedom is established. Scattering-equivalent
interactions obey asymptotic conditions beyond equality of bound-state
spectra~\cite{Ekstein1960,CoesterPolyzou1982}. Equivalent pair interactions and
three-body interactions are related through consistent cluster
transformations~\cite{PolyzouGlockle1990}, and field redefinitions redistribute
off-shell and many-body contributions~\cite{FurnstahlHammerTirfessa2001}.
Finite-rank transformations, variational operator functionals, and the induced
three-body interaction were explicitly developed in Ref.~\cite{Polyzou2010}.
Related nuclear applications fit many-body energies using phase-equivalent
pair interactions~\cite{Shirokov2007JISP16,Shirokov2012DETPET}.
Similarity-renormalization calculations also demonstrate the importance of
retaining induced many-body operators~\cite{BognerFurnstahlPerry2008,JurgensonNavratilFurnstahl2009}.

There is a direct quark-model precedent. Sengbusch and Polyzou constructed a
different confining interaction with the same meson spectrum and a consistently
changed current representation~\cite{SengbuschPolyzou2004}. A complementary
reduction of virtual $qsQg$ configurations induces both mass corrections and
effective current operators~\cite{KimNamBaek2026Hybrid}, providing a concrete
example of the correlated changes produced by eliminating additional
constituent configurations. Representation freedom and meson isospectrality
provide the starting point here. Our question
concerns their quantitative consequences for several baryon flavor and spin
sectors. We compare ground-state reproduction, predictions for excluded
excitation gaps, and the connected operator remaining under a finite embedded
transformation in a common model.

Ground-state fits, excitation gaps, and connected matrix elements constrain
different aspects of a proposed replacement. The mass fit measures how well
selected eigenvalues are reproduced, whereas excluded excitation gaps test
the resulting interaction beyond its calibration data. Evaluating the
connected operator further specifies the transformation linking the two
representations and the pair subtraction used to compare them. We carry out
all three comparisons in a common constituent S-wave space, using bounded
finite-rank pair generators with shared flavor and spin dependence.

\section{Reference model and calibration}
\subsection{Local pair interaction}
After subtracting rest masses and the center-of-mass kinetic energy, write
\begin{equation}
 H=H_2+W,\qquad H_2=T+\sum_{i<j}v_{ij}.
 \label{eq:H}
\end{equation}
The spectator is unchanged by $v_{ij}$. We use the local pair form of
Ref.~\cite{Baek2026ThreeBody}, with an independent meson calibration.
In color-singlet meson normalization the interaction is
\begin{align}
 v(r)={}&-\frac{\kappa}{r}+\sigma r-D\nonumber\\
 &+\frac{\alpha\gamma\hc}{m_i m_j r}
 e^{-(\alpha r/\hc)^2}\bm\sigma_i\cdot\bm\sigma_j.
 \label{eq:pairlocal}
\end{align}
The pair Hamiltonian is $h_c=t+cv$, with $c=1$ for a singlet $q\bar q$ and
$c=1/2$ for an antitriplet $qq$. We impose $m_u=m_d=m_q$. The regulator is
mass independent, corresponding to $\beta=\delta=0$ in a more general
mass-dependent smearing prescription. We determine all eight remaining
parameters from an independent meson fit.

The calibration uses the fixed benchmark masses of
$\pi,\rho,K,K^*,D,D^*,\eta_c(1S),J/\psi,\eta_c(2S)$ and $\psi(2S)$.
The two radial charmonium inputs are part of the meson calibration.
The objective is an unweighted squared mass residual, so the quoted RMSE
measures model error rather than experimental statistical significance.
Six starts converge to the same reported meson minimum, $4.64781$ MeV.
The resulting parameters are listed in Table~\ref{tab:local} and frozen
throughout the baryon comparisons.
\begin{table}[b]
\caption{Independent S-wave meson calibration. Energies are in MeV and
distances in fm. Fixed $\beta=\delta=0$ are excluded from the parameter count.}
\label{tab:local}
\begin{ruledtabular}
\begin{tabular}{lr@{\qquad}lr}
$\kappa$ (MeV fm)&125.26538&$\sigma$ (MeV/fm)&849.10321\\
$D$&912.39963&$m_q$&496.38496\\
$m_s$&693.17107&$m_c$&1933.91436\\
$\alpha$&1384.65773&$\gamma$&254.70885
\end{tabular}
\end{ruledtabular}
\end{table}

\subsection{Connected three-quark reference}
We take the color--spin operator form of Ref.~\cite{Baek2026ThreeBody} and fit
its strengths independently. In the $qqq$ color singlet it reads
\begin{align}
 W_f=\frac{128}{9}f\bigg[&A\sum_k\frac1{m_k}
 +B\sum_{i<j}\frac{\bm\sigma_i\cdot\bm\sigma_j}
 {m_1m_2m_3m_k^2}\nonumber\\
 &+C\sum_{i<j}\frac{\bm\sigma_i\cdot\bm\sigma_j}{m_i m_j}
 \left(\frac1{m_i}+\frac1{m_j}\right)\bigg],
 \label{eq:W}
\end{align}
where $k$ is the remaining constituent in each pair sum. The dimensions of
$(A,B,C)$ are $(\MeV^2,\MeV^6,\MeV^4)$. With
$\nu_{ij}=2m_i m_j/(m_i+m_j)$, the profiles adopted from
Ref.~\cite{Baek2026ThreeBody} are
\begin{align}
 f_G&=\exp\left[-\sum_{i<j}\frac{\nu_{ij}^2r_{ij}^2}{8(\hc)^2}\right],
 \label{eq:fG}\\
 f_E&=\exp\left[-\sum_{i<j}\frac{\nu_{ij}r_{ij}}{\hc}\right].
 \label{eq:fE}
\end{align}
Here $f_E$ is a pure exponential in the three pair distances. The central
C--C component is $W_{CC}=(128A/9)f\sum_k1/m_k$.

Nine benchmark baryon levels define the calibration set:
$N,\Delta,\Sigma,\Lambda,\Sigma_c,\Lambda_c,\Xi_{cc},\Sigma^*$ and
$\Sigma_c^*$. Eight excluded levels are
$\Xi,\Xi^*,\Xi_c,\Xi_c',\Xi_c^*,\Omega,\Omega_c$ and $\Omega_c^*$.
The $\Xi_c$ and $\Xi_c'$ levels are the lowest two roots of one Hamiltonian;
there are sixteen independent operator sectors. This split is fixed before
each fit.

Our primary three-quark reference is the unrestricted exponential fit, which
converges to $(A,B,C)\sim(-,-,+)$, with
\begin{align}
 A&=-6.70048205\times10^4\ \MeV^2,\nonumber\\
 B&=-8.29462902\times10^{14}\ \MeV^6,\nonumber\\
 C&=+3.01884095\times10^9\ \MeV^4.
 \label{eq:ABC}
\end{align}
Gaussian and exponential fits constrained to $(-,+,-)$ serve as additional
reference constructions. Their sign restriction is identified in the results.
Each Hamiltonian is rediagonalized during the fit; wave-function readjustment
is included.

\section{Meson-isospectral pair family}
Choose normalized S-wave packets
\begin{equation}
 \begin{split}
 p_a(\bm r)&=\left(\frac{2}{\pi R_a^2}\right)^{3/4}e^{-r^2/R_a^2},\\
 (R_1,R_2)&=(0.2,0.6)\ \fm.
 \end{split}
\end{equation}
Put $f=p_1$, $g=(p_2-\langle p_1|p_2\rangle p_1)/
\sqrt{1-|\langle p_1|p_2\rangle|^2}$ and
\begin{equation}
 a_{ij}=\sum_{s=0,1}\theta_s(\mu_{ij})P_{ij}^{(s)}
 (\ket f\bra g-\ket g\bra f),
 \label{eq:a}
\end{equation}
where $\mu_{ij}=m_i m_j/(m_i+m_j)$ and $P^{(s)}$ projects on pair spin.
The operator is anti-Hermitian, bounded, rotationally invariant, and
spectator independent. Its exact exponential is a two-dimensional real
rotation on each packet subspace and the identity on its complement.

The angle depends only on pair spin and reduced mass:
\begin{align}
 z_{ij}&=\frac{\mu_{ij}}{\mu_{ij}+500\ \MeV}
       -\frac{m_q/2}{m_q/2+500\ \MeV},\\
 \theta_s(z)&=a_s+b_s z+c_s z^2.
 \label{eq:angles}
\end{align}
Three parameters impose $b_0=b_1$ and $c_0=c_1=0$; four permit separate
slopes; six permit separate curvatures. The coefficients are shared by all
baryons. The three-coefficient mass fits use coefficient boxes
$|a_s|\le1$ and $|b|\le3$. The four- and six-coefficient mass fits also impose
$|\theta_s|\le1$ for the six flavor pairs. The finite operator fits use
$|a_s|\le1.5$ and $|b|\le4$ in the three-coefficient case; their reported
solutions lie inside the effective-angle range of the mass fits.
We assess the six-parameter curvature bound by relaxing its coefficient
box, and examine range sensitivity with two additional fixed radius pairs.

To vary the radial flexibility independently of this angle-law comparison,
we also use nested packet spaces of dimension $d=2,3,4,5$, adding radii
$R_3=\sqrt{0.12}$ fm, $R_4=1.0$ fm, and $R_5=1.5$ fm to the original pair.
The fifth packet extends the long-distance span; all radii are fixed before
their respective fits.
Cholesky orthogonalization of the Gaussian overlap matrix gives
$q_\alpha=\sum_\beta p_\beta C_{\beta\alpha}$ with $C^\mathsf{T}SC=I$;
its ordered construction preserves each smaller packet space inside the next.
We allow every real rotation plane in that space,
\begin{align}
 a_{ij}^{(d)}={}&\sum_{s=0,1}P_{ij}^{(s)}
 \sum_{\alpha<\beta}\theta_{s,\alpha\beta}(z_{ij})
 \bigl(\ket{q_\alpha}\bra{q_\beta}
       -\ket{q_\beta}\bra{q_\alpha}\bigr),\nonumber\\
 \theta_{s,\alpha\beta}(z)={}&a_{s,\alpha\beta}+b_{\alpha\beta}z.
 \label{eq:multipacket}
\end{align}
Each plane has two spin intercepts and one common mass slope, giving
$3$, $9$, $18$, and $30$ universal coefficients, respectively. The coefficient
boxes $|a_{s,\alpha\beta}|\le1.5$ and $|b_{\alpha\beta}|\le4$ are common
to all dimensions, and padding a smaller solution with zeros embeds it
exactly in the larger family. Here $d$ counts radial packets; the rank of a
real antisymmetric generator is even, with maximum rank two for $d=3$
and four for $d=4,5$.

For $U_{ij}=e^{a_{ij}}$, define the color-covariant extension
\begin{equation}
 v'_{ij,c}=U_{ij}(t_{ij}+cv_{ij})U_{ij}^\dagger-t_{ij}.
 \label{eq:vprime}
\end{equation}
Both $c=1$ and $c=1/2$ isolated spectra are exactly preserved. This extension
has a color-identity term:
\begin{equation}
 v'_c=cUvU^\dagger+(UtU^\dagger-t).
 \label{eq:coloridentity}
\end{equation}
This color-identity contribution enlarges the original single-Casimir
ansatz. Equation~\eqref{eq:coloridentity} includes the full change in the pair
kinetic energy when transferring the transformed meson interaction to the
$qq$ sector; direct multiplication by $1/2$ gives a different extension.
The confining isolated-pair spectra provide the isospectral constraint used
throughout this study.

We use three distinct objectives. Experimental fits use the nine benchmark
baryon masses. Direct replacement fits instead use the nine ground levels of
$H_2+W_f$ as synthetic targets. Operator fits use connected low-state matrices
defined below. Their coefficients are separately labeled. Predictions for
uncalibrated excitation gaps initially use the ground-mass fits. A second
comparison fits both the ground levels and the first gaps in the nine
calibration sectors.

For the expanded ground-mass fits at $d=2,3,4$, the nine calibration levels
provide as many or fewer constraints than the angle coefficients. We
therefore select a representative interaction by a quadratic penalty on
the generator itself,
\begin{align}
 \mathcal G_d^2&=\frac1{24}\sum_{f=1}^{6}\sum_{s=0,1}
             \|K_{fs}^{(d)}\|_F^2,\nonumber\\
 L_\lambda&=\frac19\sum_{b\in\mathrm{train}}
             (M_b^{\mathrm{pair}}-M_b^{\mathrm{ref}})^2
             +\lambda^2\mathcal G_d^2.
 \label{eq:ridge}
\end{align}
Here $K_{fs}^{(d)}$ is the real antisymmetric generator on the orthonormal
packet space for each of the six flavor pairs and two spin channels.
The penalty is invariant under orthogonal changes of packet basis, and
its normalization is kept fixed as $d$ grows. The primary scale
$\lambda=10$ MeV and sensitivity values $3$ and $30$ MeV are fixed before
examining excluded masses or excitation gaps. They set the generator penalty.
The same coefficient boxes as
in Eq.~\eqref{eq:multipacket} apply at each dimension.

For the joint spectral fit, write $\delta M_b=M_b^{\mathrm{pair}}-M_b^{\mathrm{ref}}$
and $\delta\Delta_b=\Delta_b^{\mathrm{pair}}-\Delta_b^{\mathrm{ref}}$.
We minimize
\begin{equation}
 L_\lambda^{M+\Delta}=\frac19\sum_{b\in\mathrm{train}}
 \left[(\delta M_b)^2+(\delta\Delta_b)^2\right]
 +\lambda^2\mathcal G_d^2.
 \label{eq:joint_spectrum}
\end{equation}
The two spectral errors have equal weight in MeV, with the same preset
penalty scales and coefficient bounds. These fits use nine ground masses
and nine first gaps. The eight excluded ground levels and seven excluded
gap sectors test their transfer beyond calibration. Representative
coefficients for the three-parameter ground-mass fit and the primary
four-packet joint fit are given in Appendix~\ref{app:coefficients}.

\section{Finite transformation and residual connected interaction}
\subsection{Consistent cluster subtraction}
Embed the pair generators in the three-body space and set
\begin{equation}
 \mathcal U_3=\exp\left(\sum_{i<j}a_{ij}\right).
 \label{eq:U3}
\end{equation}
The exact connected remainder associated with this representation is
\begin{align}
 W'(\theta)&=\mathcal U_3(H_2+W)\mathcal U_3^\dagger
              -H_2'(\theta),\label{eq:residual}\\
 H_2'(\theta)&=T+\sum_{i<j}v'_{ij,1/2}(\theta).
\end{align}
Here connected denotes the remainder after subtracting the one-body and
spectator-embedded pair terms specified above, as in the operational
definition of Ref.~\cite{Baek2026ThreeBody}, Sec.~4.3. This fixes the
allocation of interactions in the confining model, for which the colored
subsystems are confined. Keeping $W'$ preserves the full spectrum;
omitting it gives the pair-only approximation whose accuracy we examine.
Equation~\eqref{eq:residual} includes the complete finite exponential, the
pair kinetic changes, and commutators with $W$. Its first-order expansion
contains $W+\sum_{a\ne b}[a_a,v_b]+[\sum_a a_a,W]$, while the numerical
calculation retains all orders in the fitted angles.

Let $P_s$ contain the lowest three eigenvectors of the reference $H_s=H_{2,s}+W_s$.
The corresponding transformed states are $P_s'=\mathcal U_{3,s}P_s$.
Matrix elements between these transformed versions of the same reference
states form the aligned low block,
\begin{equation}
 B_s(\theta)=P_s'^\dagger W_s'(\theta)P_s'
 =E_s-P_s'^\dagger H_{2,s}'(\theta)P_s',
 \label{eq:block}
\end{equation}
where $E_s=P_s^\dagger H_sP_s$ is diagonal in the chosen eigenbasis and
transforms with a change of basis in that subspace. The aggregate relative
block measure is
\begin{equation}
 \epsilon_B^2(\theta)=
 \frac{\sum_s\|B_s(\theta)\|_F^2}
 {\sum_s\|P_s^\dagger W_sP_s\|_F^2}.
 \label{eq:epsB}
\end{equation}
The block thus compares the same physical low-state subspace after the
transformation and is invariant under basis rotations within that
subspace. It measures matrix elements among the selected states, while
transitions to all other states enter through the action $W_s'P_s'$.
We call this the low-input action: the residual operator acts on the three
selected low states, and every retained output direction contributes to its
norm. In a finite Ritz space $\Pi_n$, its relative measure is
\begin{equation}
 \epsilon_{A,n}^2=\frac{\sum_s\|W'_{n,s}P_s'\|_F^2}
 {\sum_s\|\Pi_n W_s P_s\|_F^2}.
 \label{eq:epsA}
\end{equation}
For the joint spectral fits we also use the two calibrated input roots
in each of the nine training sectors. This gives eighteen inputs, with
all Ritz output directions included and the denominator restricted to
the same inputs. It tests the action directly on the levels used in
the spectral calibration.
The action norm contains the low-block contribution together with the
component orthogonal to $P_s'$. Their unnormalized squared norms add,
whereas Eqs.~\eqref{eq:epsB} and \eqref{eq:epsA} use different source
normalizations. The inclusion of higher-energy output components makes
the action more sensitive to short distances, so we examine its radial
convergence separately.

To compare the spectral and block objectives using one coefficient family,
we also minimize
\begin{equation}
 J_\alpha=\alpha\frac{S_M(\theta)}{S_M(0)}
 +(1-\alpha)\epsilon_{B,\mathrm{train}}^2(\theta),
 \label{eq:joint}
\end{equation}
where $S_M$ is the nine-level synthetic ground-mass SSE. The three interior
weights $\alpha=1/4,1/2,3/4$ use both endpoint solutions as starts. The
resulting stationary points show how the fitted spectrum changes as more
weight is assigned to suppressing the connected block.

\subsection{Spectral minimum for an unrestricted unitary}
A useful comparison is the smallest action obtainable if any unitary matrix
were allowed at a fixed pair-only Hamiltonian. This spectral minimum, or
spectral floor, depends solely on how closely its eigenvalues can be matched
to the selected reference energies. For a finite pair-only matrix $K$ with
eigenvalues $\lambda_l$ and reference eigenvalues $\varepsilon_j$, it obeys
\begin{equation}
 \min_U\sum_{j=1}^k\|(UHU^\dagger-K)UP_j\|^2
 =\min_{\pi\ {\rm injective}}\sum_{j=1}^k
 (\lambda_{\pi(j)}-\varepsilon_j)^2.
 \label{eq:assignment}
\end{equation}
Indeed, in the $K$ eigenbasis the objective is
$\sum_{lj}|V_{lj}|^2(\lambda_l-\varepsilon_j)^2$ with orthonormal columns
$V$. A rectangular assignment is its minimum and can be realized by choosing
columns of the $K$ eigenbasis. The assignment therefore gives a lower
comparison value for the action of the pair-generated transformation at
that same $K$, with arbitrary matrix unitaries replacing the pair and
cluster constraints.

\section{Numerical implementation and convergence}
The variational basis contains the three Jacobi rearrangements of
$e^{-a x_{ij}^2-b y_{k,(ij)}^2}$, with all retained total-spin channels and
the required Pauli projectors~\cite{HiyamaKinoKamimura2003}. Pair-S and
spectator-S products define the total-$L=0$ orbital space used here.
Geometric radial ranges span
$0.04$--$2$ fm, scaled by the fresh constituent masses. Mass, block, and
weighted-objective comparisons initially use $n=9$ ranges, with their
original action fits at $n=11$. The principal packet-dimension action
comparison is refitted at $n=13$. Regularized ground-mass fits and excitation
predictions use $n=11$, followed by fixed-coefficient $n=13$ and $15$
gap checks. The action refinements use larger sector-specific radial spaces,
as specified with their results. Overlap eigenvalues below $10^{-10}$ of
the largest are removed.

Gaussian local and packet-kernel matrix elements are analytic. The
exponential three-distance profile uses the same converged integral
representation at quadrature orders $24$ and $32$ for $n=9$ and $11$.
Thus that comparison refines the basis and quadrature together.
The eight physical meson parameters are fixed in every comparison.

For the finite transformation we exponentiate the exact compressed bounded
generator $A_n=\Pi_n(\sum a_{ij})\Pi_n$. Equation~\eqref{eq:vprime} is
evaluated with the continuous pair kernels before compression. Therefore a
finite-$n$ discrepancy can arise between the compressed global exponential
and the continuous pair subtraction, even when all algebraic identities hold.
We quantify it with an auxiliary unsymmetrized single-pair calculation.
For one active pair and a free spectator the continuum connected remainder
vanishes exactly. Its finite-basis remainder is a projection error.
The unsymmetrized auxiliary space is essential because an individual pair
generator need not preserve the final three-body permutation sector.

The radial-complete closed linear span of rearranged S-wave products is preserved by
the packet generators: integrating one correlated Gaussian against a pair
packet produces an active-pair packet times a spectator S-wave Gaussian with
a new width. This closure supports a radial convergence study of the bounded unitary
in the chosen model space. Appendix~\ref{app:projection} gives the explicit action
and the corresponding projection-error calculation.

Independent checks compare the continuous pair kernel with an isolated-pair
matrix transformation and verify the finite three-body exponential by a
separate implementation. Pair spectral errors are below $10^{-6}$ MeV in
the dedicated packet-enriched check; finite-Ritz unitary spectral errors are
below $10^{-9}$ MeV. Reconstruction and derivative checks are provided with
the code.

\section{Results}
Unless explicitly labeled as fits to experimental benchmark masses,
the spectral comparisons below use the stated three-quark Hamiltonian
$H_2+W_f$ as their reference. Gaps refer to its ordered total-$L=0$
constituent levels; the mass-only and joint fits differ in which gaps enter
the calibration. Standard action norms use the lowest three reference inputs per
sector and every output direction in the chosen Ritz space. Radial
refinement changes both the transformation and the states, so the
finite-basis action has no general variational upper-bound property.
The one-pair projection calculation supplies an auxiliary error scale,
while separate radial refinements test the interacting three-pair problem.
The conclusions concern the specified packet spaces and locally optimized
transformations.
\subsection{Experimental masses and synthetic replacement}
\begin{table*}[tp]
\caption{Ground-state RMSEs in MeV at fixed $n=9$ fit coefficients evaluated at $n=11$.
The first block uses benchmark experimental masses; the second uses the primary
exponential three-quark reference as a synthetic target. All eight local meson
parameters are frozen. Sign constraints apply only to the explicitly labeled
three-quark references. $U_p$ denotes $p$ shared angle coefficients.}
\label{tab:ground}\begin{ruledtabular}
\begin{tabular}{llrrr}
Objective & Model & Additional parameters & Calibration 9 & Excluded 8\\
\hline
Benchmark masses & Local & 0 & 125.41 & 60.16\\
Benchmark masses & $W_G\;(-,+,-)$ & 3 & 31.52 & 27.98\\
Benchmark masses & $W_E\;(-,+,-)$ & 3 & 24.77 & 20.78\\
Benchmark masses & $W_E\;(-,-,+)$ & 3 & 8.05 & 11.46\\
Benchmark masses & $U_3$ & 3 & 21.56 & 17.72\\
Benchmark masses & $U_4$ & 4 & 14.55 & 20.49\\
Benchmark masses & $U_6$ & 6 & 13.87 & 18.71\\
\hline
$W_E\;(-,-,+)$ masses & Local & 0 & 125.00 & 54.91\\
$W_E\;(-,-,+)$ masses & $U_3$ & 3 & 21.19 & 9.44\\
$W_E\;(-,-,+)$ masses & $U_4$ & 4 & 13.63 & 15.48\\
$W_E\;(-,-,+)$ masses & $U_6$ & 6 & 12.01 & 13.18\\
\end{tabular}\end{ruledtabular}\end{table*}
\begin{table}[tbp]
\caption{Range and coefficient dependence of replacement fits to the best exponential reference.
Values are $n=11$ ground-state RMSEs in MeV. Primary radii are $(0.2,0.6)$ fm;
shorter and longer packets use $(0.15,0.45)$ and $(0.3,0.9)$ fm.
Radii are fixed within each fit. The relaxed six-coefficient fit has no active
coefficient or effective-angle constraint.}
\label{tab:controls}\begin{ruledtabular}
\begin{tabular}{lrr}Coefficients / prescription & Train 9 & Test 8\\\hline
3, primary & 21.19 & 9.44\\
4, primary & 13.63 & 15.48\\
6, primary & 12.01 & 13.18\\
6, relaxed guard & 11.97 & 14.56\\
3, shorter packets & 16.60 & 9.74\\
3, longer packets & 26.10 & 12.85\\
\end{tabular}\end{ruledtabular}\end{table}
The unrestricted exponential three-quark fit gives training/test RMSEs of
$8.05/11.46$ MeV with three additional coefficients. The three-parameter
pair family fitted directly to experiment gives $21.56/17.72$ MeV.
Four and six pair coefficients improve the training residual while their
excluded-state residuals are about $19$--$20$ MeV. These results compare
the finite pair families with the parameter counts and color extensions
specified in Sec.~III.

The synthetic replacement fit isolates the task of reproducing a specified
three-quark reference. For the primary exponential $(-,-,+)$ reference,
three pair coefficients reduce the local baseline discrepancy from
$125.00/54.91$ MeV to $\DirectThreeTrain/\DirectThreeTest$ MeV.
Increasing the number of coefficients improves the training objective but
does not improve the excluded-state score monotonically.
The weaker constrained $(-,+,-)$ reference is easier to reproduce; its
six-coefficient test residual is $4.11$ MeV. 

Table~\ref{tab:controls} shows the effect of relaxing the curvature bound
and changing the fixed packet radii. These variations distinguish the
sensitivity to the coefficient bounds from the sensitivity to the radial
support of the pair generator.

\begin{figure*}[tp]
\includegraphics[width=0.95\textwidth]{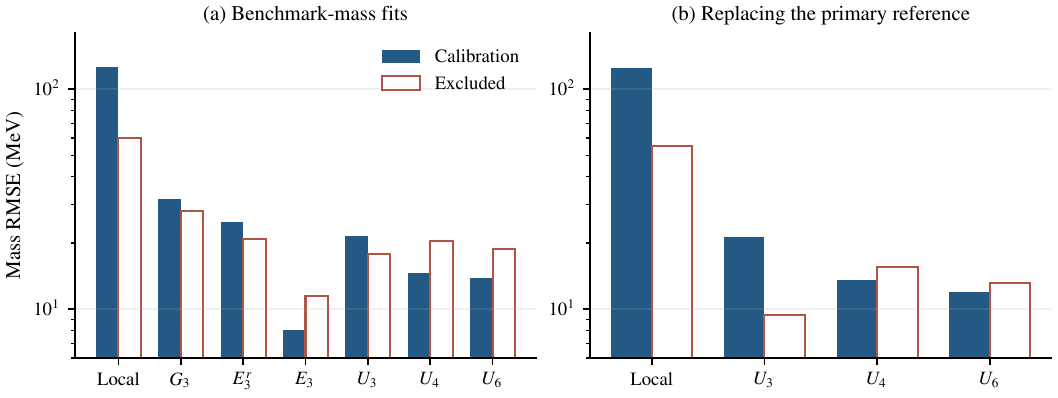}
\caption{Ground-state residuals in MeV at $n=11$. Left: fits to benchmark
experimental masses. Right: fits to the primary exponential three-quark
reference masses. Filled and open bars denote the nine calibration and eight
excluded levels. $G_3$ and $E_3^r$ denote the Gaussian and exponential
$(-,+,-)$ reference fits, $E_3$ the unrestricted exponential reference, and $U_p$
the $p$-coefficient pair family. Different objectives determine different
pair coefficients.}
\label{fig:ground}
\end{figure*}

\subsection{Regularized ground fits and uncalibrated excitation gaps}
For each independent sector the first uncalibrated gap is measured from its
lowest root. This uses root 1 in fifteen sectors and root 2 in the shared
$\Xi_c/\Xi_c'$ Hamiltonian, whose two lowest roots belong to distinct ground
multiplets.

The original two-packet fits with three, four, or six angle coefficients
leave a gap RMS difference of about $26$ MeV. To examine the additional
radial freedom, we fit the ground masses at $d=2,3,4$ using
Eq.~\eqref{eq:ridge}, with all excitation energies excluded from fitting.
Table~\ref{tab:regularized} reports the primary and preset sensitivity
choices together.

\begin{table*}[tp]
\caption{Regularized ground-mass fits and uncalibrated excitation gaps
at $n=11$. The primary penalty is $\lambda=10$ MeV; $3$ and $30$ MeV are
preset sensitivity choices. Ground errors use nine calibration and eight
excluded levels. Gap RMS values use sixteen independent sectors and the
seven excluded sectors, respectively; every gap is excluded from fitting.
All energy errors are in MeV. The penalty and coefficient bounds are the
same at each dimension.}
\label{tab:regularized}\begin{ruledtabular}
\begin{tabular}{rrrrrrrr}
$\lambda$ (MeV) & $d$ & Coefficients & $\mathcal G_d$ & Ground train & test & Gap all & test\\\hline
3 & 2 & 3 & 0.159 & 21.185 & 9.309 & 26.063 & 16.920\\
3 & 3 & 9 & 0.599 & 1.952 & 19.873 & 42.166 & 37.296\\
3 & 4 & 18 & 0.415 & 0.089 & 19.168 & 89.123 & 52.138\\
\hline
10 & 2 & 3 & 0.159 & 21.185 & 9.316 & 26.058 & 16.925\\
10 & 3 & 9 & 0.362 & 3.965 & 13.525 & 59.480 & 49.372\\
10 & 4 & 18 & 0.395 & 0.956 & 15.888 & 93.806 & 56.965\\
\hline
30 & 2 & 3 & 0.159 & 21.185 & 9.376 & 26.020 & 16.970\\
30 & 3 & 9 & 0.292 & 5.742 & 9.336 & 62.489 & 59.861\\
30 & 4 & 18 & 0.261 & 5.703 & 10.896 & 108.869 & 102.337\\
\end{tabular}\end{ruledtabular}\end{table*}
At $\lambda=10$ MeV, the first-gap RMS differences are
$\RegGapTwo$, $\RegGapThree$, and $\RegGapFour$ MeV for $d=2,3,4$.
The four-packet fit reproduces the nine reference ground masses with
$\RegMassFourTrain$ MeV RMSE, while the eight excluded masses differ by
$\RegMassFourTest$ MeV. Greater radial freedom improves the calibration
spectrum while changing the predicted excitation spectrum substantially.
The penalty sensitivity gives gap RMS values of $42$--$62$ MeV at $d=3$
and $89$--$109$ MeV at $d=4$.

A stronger penalty can recover the excluded ground-mass accuracy while
preserving a large gap difference. At $\lambda=30$ MeV and $d=3$, the
excluded ground RMSE is $9.34$ MeV, close to the two-packet primary value
of $9.32$ MeV, but the gap RMS is $62.49$ MeV. The $\Xi_{cc}$ gap varies
less with packet dimension than the other fifteen sectors in this
comparison: at $\lambda=10$ MeV its differences are $-32.95$, $-32.91$,
and $-39.27$ MeV for $d=2,3,4$, respectively.

\begin{figure*}[tp]
\includegraphics[width=0.95\textwidth]{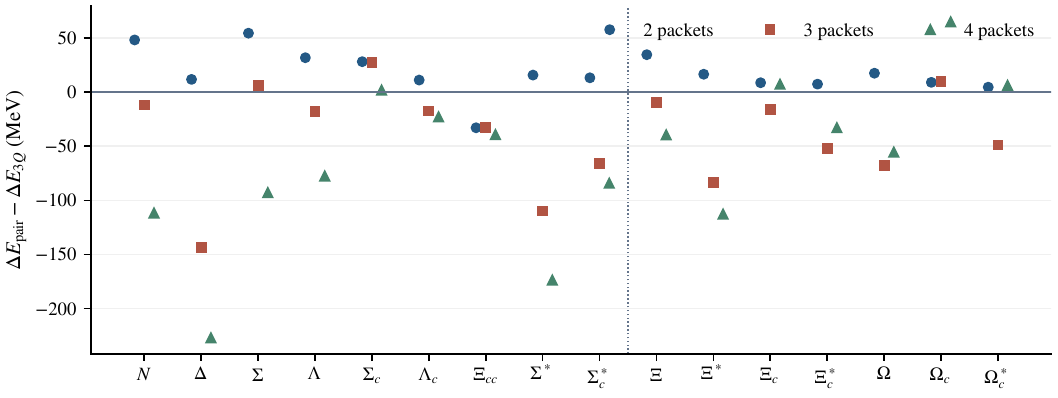}
\caption{First uncalibrated gap differences for the regularized ground-mass
fits at $\lambda=10$ MeV, evaluated at $n=11$. Packet dimension changes the
radial freedom as well as the coefficient count. The shared
$\Xi_c/\Xi_c'$ sector is counted once.}
\label{fig:gaps}
\end{figure*}

Across all nine penalty/dimension choices, fixed-coefficient
$N,\Sigma,\Omega$ checks change the pair gaps by at most
$\RegGapRefineMax$ MeV from $n=13$ to $15$; their discrepancies from the
reference change by at most $\RegGapDifferenceRefineMax$ MeV. The initial
$n=11\to13$ pair-gap shift reaches $\RegGapInitialMax$ MeV.
These representative refinements are small compared with the excitation
differences in Fig.~\ref{fig:gaps}.

\subsection{Joint calibration of ground levels and gaps}
A mass-only fit selects one regularized interaction and tests its excitation
predictions. The joint objective in Eq.~\eqref{eq:joint_spectrum} searches
directly for an interaction reproducing the ground levels and first gaps
together, with equal weight per MeV and the same generator penalty.
The primary scale is $\lambda=10$ MeV, with $3,30$ MeV as preset
four-packet sensitivity choices. The fits use $n=11$; all spectra and
actions in Table~\ref{tab:joint_spectral_action} are evaluated at $n=13$
with the coefficients fixed.
For each primary fit, the objectives from two starting points agree to
within $3\times10^{-8}$ MeV$^2$, and both solutions lie inside the
coefficient bounds.

\begin{table*}[tp]
\caption{Joint ground-mass and first-gap fits: coefficients are determined
at $n=11$ and every entry is evaluated at $n=13$. Energy errors are relative
to the three-quark reference, in MeV. Ground scores use nine calibration
and eight excluded levels; gap scores use nine calibration and seven
excluded sectors. The standard $\epsilon_A$ uses three inputs in all sixteen
sectors (48 inputs). The last column restricts the action to the two
calibrated reference roots in the nine calibration sectors (18 inputs),
with the full Ritz output space in both cases.}
\label{tab:joint_spectral_action}\begin{ruledtabular}
\begin{tabular}{rrrrrrrrr}
$\lambda$ (MeV) & $d$ & Coefficients & Ground train & test & Gap train & test &
$\epsilon_A$ all & $\epsilon_{A,\mathrm{cal}}^{(2)}$\\\hline
10 & 2 & 3 & 25.151 & 11.033 & 22.608 & 15.998 & 1.225 & 1.248\\
10 & 3 & 9 & 11.638 & 12.550 & 7.432 & 17.221 & 1.987 & 1.837\\
10 & 4 & 18 & 1.613 & 10.692 & 1.387 & 8.121 & 2.372 & 2.342\\
3 & 4 & 18 & 0.567 & 11.370 & 0.932 & 6.802 & 2.616 & 2.599\\
30 & 4 & 18 & 6.470 & 8.679 & 2.006 & 3.233 & 1.959 & 1.915\\
\end{tabular}\end{ruledtabular}\end{table*}
The four-packet primary fit gives calibration ground and gap errors of
$\JointGroundTrain$ and $\JointGapTrain$ MeV, respectively. Its excluded
ground and gap errors are $\JointGroundTest$ and $\JointGapTest$ MeV.
Thus the additional radial freedom can reproduce the selected ground
levels and first gaps together. The $\Xi_{cc}$ gap difference also changes
to $\JointXiCCGap$ MeV, resolving the persistent mass-only discrepancy
seen in Fig.~\ref{fig:gaps}. Across all five joint fits, the representative
$N,\Sigma,\Omega$ checks from $n=13$ to $15$ change the pair ground levels
and gaps by at most $\JointGroundRefineMax$ and $\JointGapRefineMax$ MeV.
The initial $n=11\to13$ gap change reaches $\JointGapInitialMax$ MeV.
The preset stronger penalty, $\lambda=30$ MeV at $d=4$, gives excluded
ground and gap errors of $\JointStrongGroundTest$ and
$\JointStrongGapTest$ MeV, with an all-sector action of
$\JointStrongActionAll$.

For the primary $d=4$, $\lambda=10$ MeV coefficients, the standard all-sector action is
$\epsilon_A=\JointActionAll$, while the unrestricted assignment minimum
at this pair Hamiltonian is $\JointActionFloor$ for the same input set.
Restricting the inputs to the two reference
roots whose energies enter the calibration gives
$\epsilon_{A,\mathrm{cal}}^{(2)}=\JointActionCal$ across the nine calibration
sectors. This restriction removes the third input from the operator test
while keeping every Ritz output direction. On these eighteen inputs, the
source and residual RMS values are $\JointCalSourceRMS$ and
$\JointCalResidualRMS$ MeV, respectively. Accurate selected eigenvalues
therefore coexist with a substantial connected action, even when its
inputs are confined to the energy levels used in the joint fit.

\begin{table}[tbp]
\caption{Action refinement for the four-packet $\lambda=10$ MeV joint fit.
The coefficients fitted at $n=11$ are held fixed. Each ratio uses the
sector's lowest $k$ reference inputs and every Ritz output direction,
with normalization by that sector's source action.}
\label{tab:joint_action_refinement}\begin{ruledtabular}
\begin{tabular}{lrrrr}
Sector & $k$ & $n=13$ & $n=15$ & $n=17$\\\hline
$N$ & 2 & 2.296069 & 2.290015 & 2.291383\\
$N$ & 3 & 2.298671 & 2.291914 & 2.293301\\
$\Sigma$ & 2 & 2.863035 & 2.848927 & 2.850063\\
$\Sigma$ & 3 & 2.867332 & 2.851433 & 2.852811\\
\end{tabular}\end{ruledtabular}\end{table}
Table~\ref{tab:joint_action_refinement} follows the action of these joint-fit
coefficients in larger $N$ and $\Sigma$ spaces. The largest change across
the two sectors and both input choices is $\JointActionNSInitialChange$
from $n=13$ to $15$, followed by $\JointActionNSLastChange$ from $15$ to
$17$. The two-input action and its three-input counterpart show the same
substantial residual in these refined spaces.

\begin{figure*}[tp]
\includegraphics[width=0.95\textwidth]{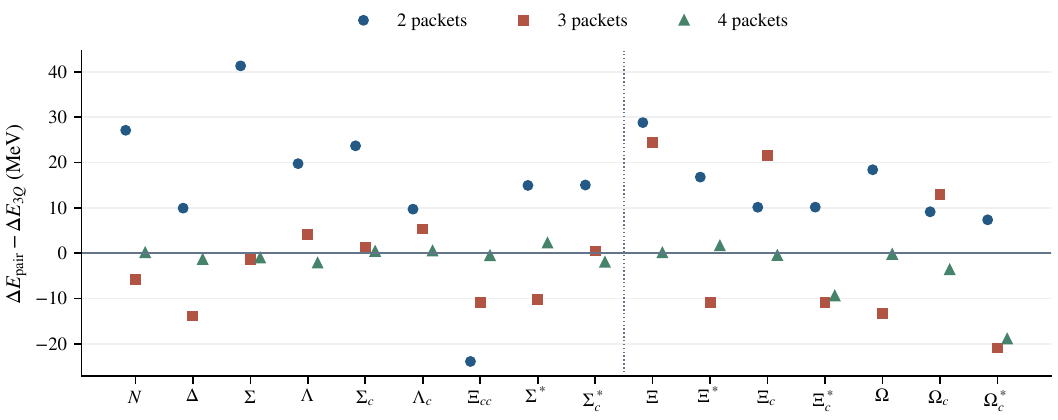}
\caption{First-gap differences for the joint ground-level and gap fits at
$\lambda=10$ MeV, evaluated at $n=13$ using the $n=11$ fitted coefficients.
The nine sectors to the left of the dotted line supply calibration gaps;
the seven to the right test excluded predictions. The shared
$\Xi_c/\Xi_c'$ sector is counted once.}
\label{fig:joint_gaps}
\end{figure*}

\subsection{Connected action refitted in a larger radial basis}
The original $n=11$ comparison illustrates the difference between fitting
a spectrum and suppressing the connected action. The ground-mass fit gives
$\epsilon_A=\MassFitAction$, compared with unity at zero rotation, while
separate block and action objectives select different coefficients.
Appendix~\ref{app:original_objectives} gives those fixed-resolution
comparisons. The principal packet-dimension results below are refitted
and evaluated at $n=13$.

\begin{table}[tbp]
\caption{Change from the original $n=11$ action fits to the principal
$n=13$ comparison across the same sixteen sectors. Column headings give
fitting/evaluation resolutions. The middle column holds the original
coefficients fixed, while the final column refits them at $n=13$.}
\label{tab:refit_shift}\begin{ruledtabular}
\begin{tabular}{rrrr}$d$ & $11/11$ & $11/13$ & $13/13$\\\hline
2 & 0.883125 & 0.862191 & 0.861055\\
3 & 0.806849 & 0.772950 & 0.752846\\
4 & 0.763777 & 0.741816 & 0.727891\\
\end{tabular}\end{ruledtabular}\end{table}
\begin{table*}[tp]
\caption{Packet-dimension action fits performed and evaluated at $n=13$.
The nine calibration sectors determine the coefficients; the other seven
operator sectors are excluded. Each row's unrestricted minimum is evaluated
at its own pair Hamiltonian. Ground RMSEs are relative to the three-quark
reference, in MeV. All rows share a source RMS of $201.32$ MeV
per input state, $\|\Pi_{13}WP\|_F/\sqrt{48}$.}
\label{tab:refined_packet}\begin{ruledtabular}
\begin{tabular}{rrrrrrrr}
$d$ & Coefficients & $\epsilon_A$ train & test & all & Unrestricted minimum & Ground train & test\\\hline
2 & 3 & 0.868961 & 0.824025 & 0.861055 & 0.278456 & 97.37 & 33.64\\
3 & 9 & 0.761550 & 0.711789 & 0.752846 & 0.199177 & 70.39 & 20.97\\
4 & 18 & 0.734291 & 0.697947 & 0.727891 & 0.165256 & 65.03 & 19.41\\
5 & 30 & 0.732283 & 0.696672 & 0.726010 & 0.165745 & 65.49 & 19.47\\
\end{tabular}\end{ruledtabular}\end{table*}
Table~\ref{tab:refit_shift} separates basis refinement at fixed coefficients
from the further change under refitting. Both lower the aggregate action
in each dimension, showing how the original $n=11$ values overstate the
residual in the larger-space calculation.

At $n=13$, the all-sector action decreases from $\RefinedActionTwo$
through $\RefinedActionThree$ and $\RefinedActionFour$ to
$\RefinedActionFive$ as the packet dimension grows from two to five.
The resulting ground errors fall through four packets and change only
slightly at five; these action-optimized spectra are less accurate than
the ground-mass fits. In absolute units the common source action RMS is
$\RefinedSourceRMS$ MeV per input state and the five-packet residual RMS
is $\RefinedFiveResidualRMS$ MeV.

\begin{table*}[tp]
\caption{Sector-resolved action refinement with the coefficients fixed
at their $n=13$ fitted values. The $N$ and $\Sigma$ sectors are evaluated
at $n=15$ and $17$, and the excluded $\Omega$ sector at $n=19$ and $21$.
Dots mark unevaluated spaces. Each sector is normalized by its own source
action.}
\label{tab:refined_radial}\begin{ruledtabular}
\begin{tabular}{lrrrrrr}Sector & $d$ & $n=13$ & $n=15$ & $n=17$ & $n=19$ & $n=21$\\\hline
$N$ & 2 & 0.886723 & 0.886929 & 0.887098 & $\cdots$ & $\cdots$\\
$N$ & 3 & 0.790532 & 0.789183 & 0.789889 & $\cdots$ & $\cdots$\\
$N$ & 4 & 0.760512 & 0.758548 & 0.759464 & $\cdots$ & $\cdots$\\
$N$ & 5 & 0.759062 & 0.757101 & 0.758043 & $\cdots$ & $\cdots$\\
\hline
$\Sigma$ & 2 & 0.872016 & 0.871472 & 0.871782 & $\cdots$ & $\cdots$\\
$\Sigma$ & 3 & 0.794069 & 0.790225 & 0.791205 & $\cdots$ & $\cdots$\\
$\Sigma$ & 4 & 0.764671 & 0.761660 & 0.762785 & $\cdots$ & $\cdots$\\
$\Sigma$ & 5 & 0.764017 & 0.761131 & 0.762296 & $\cdots$ & $\cdots$\\
\hline
$\Omega$ & 2 & 0.912513 & $\cdots$ & $\cdots$ & 0.856369 & 0.856751\\
$\Omega$ & 3 & 0.711710 & $\cdots$ & $\cdots$ & 0.660991 & 0.661236\\
$\Omega$ & 4 & 0.671157 & $\cdots$ & $\cdots$ & 0.618930 & 0.619105\\
$\Omega$ & 5 & 0.668310 & $\cdots$ & $\cdots$ & 0.619884 & 0.620116\\
\end{tabular}\end{ruledtabular}\end{table*}
The fixed-coefficient refinements in Table~\ref{tab:refined_radial} follow
individual sectors, including the more slowly converging excluded
$\Omega$ channel. With the $n=13$ fitted coefficients held fixed, the
largest $n=13\to15$ change for $N$ and $\Sigma$ is
$\RefinedNSInitialChange$, followed by $\RefinedNSLastChange$ from
$n=15$ to $17$. The largest final $n=19\to21$ change for $\Omega$ is
$\RefinedOmegaLastChange$, across the four packet dimensions.
At $n=21$, the five-packet $\Omega$ residual exceeds the four-packet
value by about $0.0010$. Appendix~\ref{app:projection} compares the
one-pair projection error with its source action in MeV.

\subsubsection{An intermediate-distance fifth packet}
The third packet already has radius $\sqrt{0.12}\simeq0.346$ fm.
We therefore add a distinct fifth radius of $0.45$ fm to examine further
freedom inside the original $0.2$--$0.6$ fm interval. This radius is fixed
before fitting, with the same thirty coefficients and coefficient boxes
as for the $1.5$ fm extension.

\begin{table*}[tp]
\caption{Fifth-packet radius comparison at $n=13$. Both fits have thirty
shared coefficients and the same first four packets, bounds, calibration
sectors, and action objective. The unrestricted minimum is evaluated at
each fit's own pair Hamiltonian. Residual action RMS and ground RMSEs are
in MeV; the common source RMS is $201.32$ MeV
across 48 input states.}
\label{tab:short_packet}\begin{ruledtabular}
\begin{tabular}{rrrrrrrr}
$R_5$ (fm) & $\epsilon_A$ train & test & all & Unrestricted minimum &
Residual RMS & Ground train & test\\\hline
1.50 & 0.732283 & 0.696672 & 0.726010 & 0.165745 & 146.16 & 65.49 & 19.47\\
0.45 & 0.715409 & 0.689233 & 0.710774 & 0.159734 & 143.09 & 63.57 & 19.08\\
\end{tabular}\end{ruledtabular}\end{table*}
The $0.45$ fm packet gives an all-sector action of $\ShortActionAll$,
compared with $\RefinedActionFive$ for the long-distance extension.
Its residual RMS is $\ShortResidualRMS$ MeV on the same 48 inputs.
The new direction changes the optimized action while leaving a substantial
connected term. With the new coefficients fixed, the excluded $\Omega$
channel changes from $\ShortOmegaAtThirteen$ at $n=13$ to
$\ShortOmegaAtNineteen$ at $n=19$ and $\ShortOmegaAtTwentyOne$ at $n=21$.
For $N$ and $\Sigma$, the largest change is $\ShortNSInitialChange$ from
$n=13$ to $15$, followed by $\ShortNSLastChange$ from $15$ to $17$.

\subsection{Comparison with unrestricted state alignment}
\begin{figure*}[tp]
\includegraphics[width=0.95\textwidth]{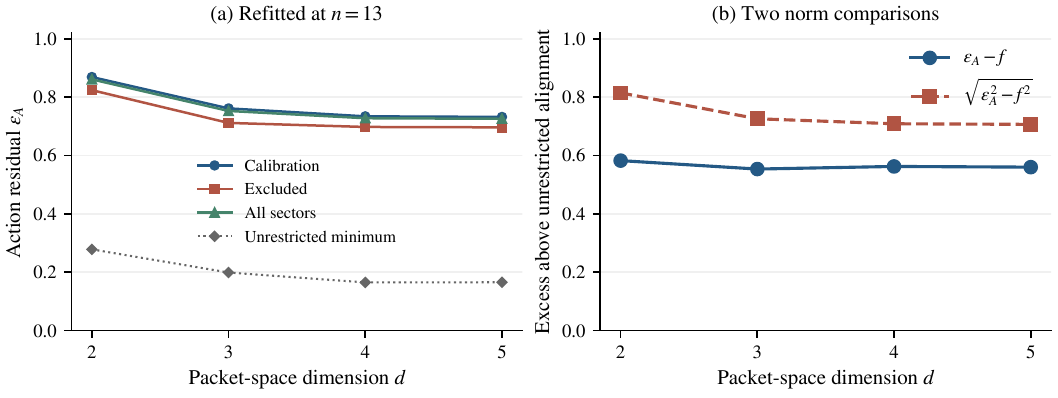}
\caption{Action fits at $n=13$ with two through five packets, using
$R_5=1.5$ fm. Left: calibration, excluded, and aggregate action residuals,
together with the unrestricted minimum at each fit's own pair Hamiltonian.
Right: linear and quadratic excesses defined in Eq.~\eqref{eq:excess},
using all sixteen sectors.}
\label{fig:packet_dimension}
\end{figure*}

For each fitted pair Hamiltonian let $f_d$ be the normalized assignment
minimum in Eq.~\eqref{eq:assignment}. Two comparisons of the achieved
action with this minimum are
\begin{equation}
 \delta_{\rm lin}=\epsilon_A-f_d,\qquad
 \delta_{\rm quad}=\sqrt{\epsilon_A^2-f_d^2}.
 \label{eq:excess}
\end{equation}
The first subtracts norms, whereas the second subtracts the corresponding
squared functionals before taking the square root. A nearly constant
linear difference can coexist with a decreasing quadratic excess, as
shown in Fig.~\ref{fig:packet_dimension}.

Each $f_d$ is evaluated at that fit's own $H_2'$. The differences in
Eq.~\eqref{eq:excess} compare the chosen state alignment with arbitrary
unitary alignment at the same matrix and reflect the combined restrictions
from packet span, shared coefficients, embedding, and the local fit.
The two choices of fifth-packet radius probe long- and intermediate-distance
directions with the same coefficient count.

\section{An explicit realization of the intermediate-configuration mechanism}
This construction explicitly realizes the central part of the $q^\ast$
intermediate-configuration mechanism in Ref.~\cite{Baek2026ThreeBody},
Eqs.~(3.20)--(3.23) and (4.24)--(4.25). There $q^\ast$ denotes a generic
configuration outside the retained valence $qqq$ space. Here $Q_k$ contains
an orthogonal excitation carried by constituent $k$, providing a concrete
choice for the intermediate line. Define
\begin{equation}
 B_k=g[\CC_{ki}u(r_{ki})+\CC_{kj}u(r_{kj})],\qquad
 \CC_{ij}=\bm\lambda_i\cdot\bm\lambda_j.
\end{equation}
For positive gaps, static second-order elimination gives
$\Delta H=-\sum_k B_k^\dagger B_k/\Delta_k$
\cite{Feshbach1958,BravyiDiVincenzoLoss2011}.
Its cross term contains
\begin{equation}
 W_k=-\frac{g^2}{\Delta_k}u(r_{ki})u(r_{kj})
       \{\CC_{ki},\CC_{kj}\}.
\end{equation}
The two pair orderings give the color relation used in
Ref.~\cite{Baek2026ThreeBody}, Eq.~(3.23),
\begin{equation}
 \{\CC_{12},\CC_{23}\}=\frac43\CC_{13}
 +2d^{abc}\lambda_1^a\lambda_2^b\lambda_3^c
 \ \longrightarrow\ \frac{128}{9}
\end{equation}
in a $qqq$ singlet. Choosing $\Delta_k=\xi m_k$ realizes the intermediate
denominator $M_k$ of Ref.~\cite{Baek2026ThreeBody} with a shared-line
inverse mass and $A=-g^2/\xi$. The gap law and transition functions are
assumed inputs; this is an existence construction for the central operator,
not a prediction of those inputs from QCD.

The same elimination also gives the pair dressing assigned to the baseline
in Ref.~\cite{Baek2026ThreeBody}, Eq.~(4.24). In this realization it is
\begin{equation}
 \delta v_{ij}=-g^2\CC_{ij}^2u(r_{ij})^2
 \left(\frac1{\Delta_i}+\frac1{\Delta_j}\right).
 \label{eq:dressing}
\end{equation}
The squared color eigenvalues are $256/9$ in a meson singlet and $64/9$ in a
$qq$ antitriplet. For charge-conjugate vertices with the same reduced
strength and gap convention, the ratio is $1/4$, which differs from the bare Casimir
ratio $1/2$. A microscopic allocation to the meson-calibrated baseline must
therefore retain the correlated self and cross terms. This energy-shifting
pair self-energy is distinct from the isospectral pair rotation in
Eq.~\eqref{eq:vprime}.

For the Gaussian three-quark profile of Ref.~\cite{Baek2026ThreeBody},
consider equal masses, let $b=m^2/[8(\hc)^2]$, and choose
$u(r)=e^{-3br^2/2}$. The normalized cross profile
\begin{equation}
 S=\frac13\sum_k e^{-3b(r_{ki}^2+r_{kj}^2)/2}
\end{equation}
agrees with $f_G$ through quadratic order in the distances and exactly for
equilateral configurations. Writing $x=r_{12}^2,y=r_{23}^2,z=r_{31}^2$,
\begin{equation}
 \frac{S}{f_G}=1+\frac{b^2}{8}[(x-y)^2+(y-z)^2+(z-x)^2]
 +O(b^3r^6).
\end{equation}
With the fresh $H_2$ lowest three states, exact continuous Gaussian-product
integrals give action errors $4.94\%$, $4.03\%$, and $4.65\%$ for
$N,\Delta$, and $\Omega$, with the amplitude and width fixed by the
construction above. This agreement provides a realization of the central
three-body cross profile together with the correlated pair dressing in
Eq.~\eqref{eq:dressing}. Establishing an isospectral replacement additionally
requires the consistent finite transformation examined in Sec.~VI.

\section{Discussion and conclusions}
The connected interaction is defined in the constituent representation of
Ref.~\cite{Baek2026ThreeBody}. Once the one-body Hamiltonian and spectator
embedding of each pair interaction have been fixed, their subtraction
defines the three-quark remainder. Its finite-range local profiles depend
jointly on several interquark distances and vanish when one constituent
is separated from the other two. The added operator therefore carries
spatial correlations beyond a sum of the specified local pair functions.
Confinement belongs to the underlying Hamiltonian; the cluster falloff
describes its additional short-range interaction.

The Polyzou--Gl\"ockle relation places this definition within a broader
freedom to redistribute few-body dynamics between equivalent
representations~\cite{PolyzouGlockle1990,Polyzou2010}. Where scattering
channels exist, the equivalence includes asymptotic conditions beyond
equality of spectra, and observable operators accompany the Hamiltonian
under the transformation~\cite{Ekstein1960,Polyzou2010}. Changing the
off-shell pair dynamics can then change the explicit many-body interaction
while preserving the full observables. Suitable representations can weaken
the three-body interaction, although its complete elimination depends on
the starting Hamiltonian and admissible transformations.

Our calculation explores a restricted part of that freedom. The general
construction allows an independent connected three-particle generator after
the transformed pair interactions have been fixed~\cite{Polyzou2010}.
Here $\mathcal U_3=\exp(\sum_{i<j}a_{ij})$ is specified entirely by the
shared pair generators. Exponentiation produces connected contributions
whose coefficients are determined by those pair inputs. The residual
therefore tests the specified pair-generated representation rather than the entire class
of equivalent Hamiltonians. In the confining setting, preservation of the
isolated spectra supplies the directly demonstrated constraint.

The exact relation is $\mathcal U_3H\mathcal U_3^\dagger=H_2' +W'$.
Together, $H_2'$ and $W'$ preserve the full dynamics; the pair-only
approximation discards $W'$. The original two-packet mass fit at $n=11$
has $\epsilon_A=\MassFitAction$, compared with unity at zero rotation.
Its close ground-state fit thus coexists with an increased norm of the
discarded connected action on the selected states.

The mass-only fits test how a regularized interaction extrapolates from
ground levels to excitations. Increasing the packet dimension improves
the calibration masses while changing those predictions, and the
regularization sensitivity shows how the selected interaction depends
on the penalty. The joint ground-level and gap fits search more directly
for an interaction that reproduces both parts of the spectrum. Their
excluded-sector predictions test transfer across flavor and spin, while
the connected action measures the relation of each fitted interaction to
the reference Hamiltonian. The four-packet joint fit gives calibration
errors of $\JointGroundTrain$ MeV for masses and $\JointGapTrain$ MeV for
gaps at $n=13$, with excluded errors of $\JointGroundTest$ and
$\JointGapTest$ MeV. Its normalized action on the eighteen calibrated
inputs is nevertheless $\JointActionCal$. The fitted eigenvalues and the
mapping of their eigenvectors by the shared pair-generated unitary carry
different information.

Optimizing the action itself gives a substantial residual in the families
examined. Each fit's unrestricted spectral minimum supplies a comparison
at the same pair Hamiltonian; its difference from the achieved action
combines restrictions from the radial span, shared coefficients, and
embedding. Neither that difference nor an extrapolation from a few packet dimensions isolates an
irreducible physical three-body component. Radial refinement and changes
of transformation family address different parts of this comparison.

These findings favor a small connected operator basis as an economical
description in the meson-calibrated model. The reference uses three fitted
strengths, while the four-packet joint pair fit uses eighteen shared
coefficients and a preset generator penalty. The couplings $A,B,C$
characterize the chosen representation, spatial profile, and calibration
data~\cite{Baek2026ThreeBody}; they can contain contributions from
eliminated gluonic, instanton, relativistic, or additional-channel dynamics
within the chosen operator channels. The eliminated-state example
explicitly realizes a common origin of pair dressing and the central
C--C structure, with a specified gap law and transition profile. Meson
and baryon masses constrain this effective description. They do not select a unique microscopic off-shell interaction.
Complete replacement is an open question for a broader transformation
class. The present comparisons show why agreement with selected masses
and suppression of the connected interaction require separate tests.

\section*{Data availability}
The data that support the findings of this article are not publicly
available upon publication because it is not technically feasible and/or
the cost of preparing, depositing, and hosting the data would be prohibitive
within the terms of this research project. The data are available from the
authors upon reasonable request.

\appendix
\section{Packet action and radial projection}
\label{app:projection}
For a correlated Gaussian written in the active pair coordinates,
$G=N\exp[-a x^2-2c\bm x\cdot\bm y-d y^2]$, integration against the ket
packet of precision $\alpha_q$ gives
\begin{equation}
 \int d^3x\,e^{-\alpha_qx^2}G
 =N\left(\frac\pi{a+\alpha_q}\right)^{3/2}
 e^{-[d-c^2/(a+\alpha_q)]y^2}.
\end{equation}
Multiplication by the output packet gives an S-wave product in the same
active pair arrangement. The exact generator action on the input wave
functions can therefore be retained as an enlarged Gaussian expansion.
Its overlap Gram yields both $\|AP\|$ and
$\|(I-\Pi_n)AP\|$, including output outside the finite basis.
This Gram calculation measures the part of the generator action omitted
by projection. Its effect on the finite-$U$ residual further depends on
the energy weighting from the Hamiltonian.

\begin{table*}[tp]
\caption{Auxiliary single-pair projection benchmark at $n=13$, using the
exponential block-fit coefficients $\theta_B$ and the lowest three $H_2+W_E$
input states. Source and projection-error RMS values are Frobenius action
norms divided by $\sqrt{3}$, in MeV; the source is $\Pi_{13}WP$ in the
physical reference space. The auxiliary continuum remainder is zero, and
the ratio compares its projection error with the source scale.}
\label{tab:projection}\begin{ruledtabular}
\begin{tabular}{lrrrr}Sector & Source RMS & Projection RMS & Ratio (\%) & Block operator norm (MeV)\\\hline
$N$ & 422.36 & 8.46 & 2.00 & 0.00191\\
$\Sigma$ & 247.08 & 19.21 & 7.77 & 0.00673\\
$\Xi_c$ & 155.61 & 19.62 & 12.61 & 0.01008\\
\end{tabular}\end{ruledtabular}\end{table*}
The auxiliary one-pair example combines a free spectator with a pair
Hamiltonian, so its exact continuum subtraction vanishes and the computed
remainder measures compression error. Low blocks and the full compressed
action converge at different rates, as shown by their separate
fixed-coefficient radial refinements. The one-pair benchmark isolates
projection error in a setting with an exact zero remainder, while the
three-pair refinements assess convergence of the interacting calculation.

\section{Benchmark inputs and state-resolved ground spectra}
\begin{table*}[tp]
\caption{Fixed S-wave meson benchmark and independent local fit, in MeV.
Both $2S$ charmonium inputs belong to the calibration.}
\label{tab:mesondata}\begin{ruledtabular}\begin{tabular}{lrrlrr}
Meson & Benchmark & Local fit & Meson & Benchmark & Local fit\\\hline
$\pi$ & 139.570 & 141.355 & $D^*$ & 2010.300 & 2006.113\\
$\rho$ & 775.110 & 768.390 & $\eta_c(1S)$ & 2983.600 & 2985.197\\
$K$ & 493.680 & 488.959 & $J/\psi$ & 3096.900 & 3094.047\\
$K^*$ & 891.660 & 900.906 & $\eta_c(2S)$ & 3637.800 & 3637.998\\
$D$ & 1864.800 & 1870.425 & $\psi(2S)$ & 3686.097 & 3686.126\\
\end{tabular}\end{ruledtabular}\end{table*}
\begin{table*}[tp]
\caption{Ground masses at $n=11$, in MeV. T and V mark the training and
excluded sets. $W_E$ is fitted to benchmark masses; $U_3$ and $U_6$ are
separately fitted to its nine training levels.}
\label{tab:baryondata}\begin{ruledtabular}\begin{tabular}{lcrrrr}
Baryon & Set & Benchmark & $H_2+W_E$ & $H'_2(\theta_{M,3})$ & $H'_2(\theta_{M,6})$\\\hline
$N$ & T & 938.272 & 934.895 & 941.335 & 926.622\\
$\Delta$ & T & 1232.000 & 1231.464 & 1233.274 & 1251.104\\
$\Sigma$ & T & 1192.642 & 1204.067 & 1170.147 & 1194.602\\
$\Lambda$ & T & 1115.683 & 1106.529 & 1134.225 & 1119.710\\
$\Sigma_c$ & T & 2452.650 & 2450.459 & 2416.837 & 2444.698\\
$\Lambda_c$ & T & 2286.460 & 2298.943 & 2324.144 & 2303.784\\
$\Xi_{cc}$ & T & 3619.970 & 3631.833 & 3634.962 & 3635.802\\
$\Sigma^*$ & T & 1383.800 & 1384.178 & 1376.258 & 1384.481\\
$\Sigma_c^*$ & T & 2518.480 & 2511.051 & 2495.511 & 2488.522\\
$\Xi$ & V & 1321.000 & 1329.636 & 1333.711 & 1334.781\\
$\Xi^*$ & V & 1535.000 & 1527.556 & 1517.188 & 1517.116\\
$\Xi_c$ & V & 2469.000 & 2494.495 & 2500.676 & 2488.465\\
$\Xi_c^{\prime}$ & V & 2579.000 & 2578.364 & 2559.187 & 2566.341\\
$\Xi_c^*$ & V & 2646.000 & 2636.969 & 2627.509 & 2615.509\\
$\Omega$ & V & 1672.000 & 1663.468 & 1655.143 & 1648.802\\
$\Omega_c$ & V & 2695.000 & 2701.163 & 2696.431 & 2688.878\\
$\Omega_c^*$ & V & 2766.000 & 2757.146 & 2755.869 & 2741.408\\
\end{tabular}\end{ruledtabular}\end{table*}
Tables~\ref{tab:mesondata} and \ref{tab:baryondata} specify the fixed
calibration values and ground spectra. The $\Xi_c/\Xi_c'$ pair occupies two
roots of one Hamiltonian; their individual benchmark errors enter the
eight-level excluded mass score, while operator and gap aggregates count
the common sector once.

\section{A spatially unresolved color sum rule}
In the uniform-vertex limit $u\to1$, color neutrality gives
$\sum_{j\ne k}\CC_{kj}=-16/3$ on the singlet. Therefore the complete static
correction is an additive constituent dressing,
\begin{equation}
 \Delta H=-\frac{256g^2}{9}\sum_k\frac1{\Delta_k}.
\end{equation}
For a meson pair dressing transferred naively by $1/2$, restoring the
microscopic $qq$ squared-color factor requires
\begin{equation}
 \delta V_{\rm transfer}=\frac{64g^2}{9}
 \sum_{i<j}\left(\frac1{\Delta_i}+\frac1{\Delta_j}\right).
\end{equation}
It cancels the isolated C--C cross term in this spatially uniform limit.
The cancellation exhibits how correlated baseline dressing changes the
allocation between pair and connected terms in the explicit eliminated-state
model. For finite profiles, the same allocation involves the spatially dependent
self and cross terms derived in Sec.~VII.

\section{Original fixed-resolution objective comparisons}
\label{app:original_objectives}
Table~\ref{tab:objectives} retains the original two-packet comparisons at $n=11$.
They isolate differences among the mass, block, action, and weighted
objectives at that resolution. The principal action and packet-dimension
comparison in Sec.~VI instead uses the $n=13$ refits.
\begin{table*}[tp]
\caption{Different three-coefficient objectives for the primary exponential
reference, all evaluated at $n=11$. Ground errors are synthetic RMSEs in MeV;
the block and action measures use their respective source norms. The last
column gives the smallest action allowed by arbitrary finite-matrix unitaries
at the corresponding fixed pair-only Hamiltonian [Eq.~\eqref{eq:assignment}].}
\label{tab:objectives}\begin{ruledtabular}
\begin{tabular}{lrrrrrr}
Objective & Fit $n$ & Ground train 9 & Ground test 8 & $\epsilon_B$ & $\epsilon_A$ & Unrestricted minimum\\\hline
Mass fit $\theta_M$ & 9 & 21.19 & 9.44 & 0.361 & 1.360 & 0.096\\
Block fit $\theta_B$ & 9 & 62.97 & 24.03 & 0.294 & 1.708 & 0.156\\
Action fit $\theta_A$ & 11 & 100.03 & 35.73 & 0.785 & 0.883 & 0.286\\
\end{tabular}\end{ruledtabular}\end{table*}
The action objective is fitted directly at $n=11$ from two starts, with both
solutions reaching the same reported basin. The remaining relative action
is $0.8824$ on the training sectors and $0.8863$ on the excluded sectors.
At this point the synthetic ground-mass RMSEs are $100.03/35.73$ MeV.
A six-coefficient action fit in the same packet family gives training, excluded, and aggregate action residuals of $0.881255$, $0.907930$, and $0.886111$ at $n=11$. Both starts reach the reported basin and no coefficient bound is active. The extra mass-angle freedom slightly improves the training objective and worsens transfer to excluded sectors. This comparison varies the angle law at fixed packet dimension and ranges.
The intermediate weighted fits in Eq.~\eqref{eq:joint} give ground-mass
RMSEs of $31.12/15.72$, $23.99/12.46$, and $21.74/10.61$ MeV at
$\alpha=0.25,0.50,0.75$, respectively, while their all-sector block
residuals increase from $0.318$ through $0.337$ to $0.351$.

\clearpage
\onecolumngrid
\section{Representative pair-rotation coefficients}
\label{app:coefficients}

Table~\ref{tab:representativecoefficients} specifies the primary four-packet
joint fit and the original two-packet ground-mass fit. The former is fitted
at $n=11$ with $\lambda=10$ MeV, and the latter at $n=9$ without a generator
penalty. The fixed coefficients are used at the evaluation resolutions
specified in the main text.

The ordered radii are $(0.2,0.6,\sqrt{0.12},1.0)$ fm for $d=4$ and
$(0.2,0.6)$ fm for $d=2$. Using the positive normalized Gaussian packets
of Sec.~III, their overlap is
$S_{\alpha\beta}=[2R_\alpha R_\beta/(R_\alpha^2+R_\beta^2)]^{3/2}$.
We factor $S=LL^{\mathsf T}$ with a lower-triangular $L$ having positive
diagonal entries, set $C=L^{-\mathsf T}$, and form
$q_\alpha=\sum_\beta p_\beta C_{\beta\alpha}$. The radii keep the displayed
order throughout this construction.

Each table row gives $(a_{0,\alpha\beta},a_{1,\alpha\beta},b_{\alpha\beta})$
in Eq.~\eqref{eq:multipacket}. The generator has upper-triangle entries
$(K_s)_{\alpha\beta}=a_{s,\alpha\beta}+b_{\alpha\beta}z$ and lower-triangle
entries $(K_s)_{\beta\alpha}=-(K_s)_{\alpha\beta}$, with $U_s=e^{K_s}$.
Here $s=0,1$ denotes singlet and triplet pair spin, respectively, and
\begin{equation}
 \begin{aligned}
 z&=\frac{\mu_{ij}}{\mu_{ij}+500\,\MeV}
   -\frac{m_q/2}{m_q/2+500\,\MeV},\\
 \mu_{ij}&=\frac{m_i m_j}{m_i+m_j}.
 \end{aligned}
\end{equation}
All coefficients are dimensionless. The basis generator
$|q_\alpha\rangle\langle q_\beta|-|q_\beta\rangle\langle q_\alpha|$
has matrix entry $+1$ at $(\alpha,\beta)$, fixing the rotation sign.
The table gives twelve decimal places.

\begin{table}[!htb]
\caption{Representative universal pair-rotation coefficients. Each plane
has two spin intercepts and one shared mass slope. Plane labels follow
the ordered Cholesky basis.}
\label{tab:representativecoefficients}
\begin{ruledtabular}
\begin{tabular}{llrrr}
Fit & Plane $(\alpha,\beta)$ & $a_0$ & $a_1$ & $b$\\\hline
Joint, $d=4$, $\lambda=10$ MeV & $(1,2)$ & $0.329880749453$ & $0.234892613387$ & $-1.331024863244$\\
 & $(1,3)$ & $-0.121231241450$ & $0.219541978534$ & $-2.081593970295$\\
 & $(2,3)$ & $-0.464611658295$ & $-0.046252184320$ & $1.325969808123$\\
 & $(1,4)$ & $-0.125316950705$ & $0.001130973923$ & $-0.297890237969$\\
 & $(2,4)$ & $-0.145788303407$ & $0.059149074520$ & $-0.154915818585$\\
 & $(3,4)$ & $-0.124094162428$ & $-0.023648949482$ & $0.014152220849$\\
\hline
Ground, $d=2$, $\bm\theta_{M,3}$ & $(1,2)$ & $0.243109368973$ & $0.230202297318$ & $-1.228243466038$\\
\end{tabular}
\end{ruledtabular}
\end{table}
\clearpage
\twocolumngrid
\bibliography{references}

\begin{thebibliography}{23}%
\makeatletter
\providecommand \@ifxundefined [1]{%
 \@ifx{#1\undefined}
}%
\providecommand \@ifnum [1]{%
 \ifnum #1\expandafter \@firstoftwo
 \else \expandafter \@secondoftwo
 \fi
}%
\providecommand \@ifx [1]{%
 \ifx #1\expandafter \@firstoftwo
 \else \expandafter \@secondoftwo
 \fi
}%
\providecommand \natexlab [1]{#1}%
\providecommand \enquote  [1]{``#1''}%
\providecommand \bibnamefont  [1]{#1}%
\providecommand \bibfnamefont [1]{#1}%
\providecommand \citenamefont [1]{#1}%
\providecommand \href@noop [0]{\@secondoftwo}%
\providecommand \href [0]{\begingroup \@sanitize@url \@href}%
\providecommand \@href[1]{\@@startlink{#1}\@@href}%
\providecommand \@@href[1]{\endgroup#1\@@endlink}%
\providecommand \@sanitize@url [0]{\catcode `\\12\catcode `\$12\catcode
  `\&12\catcode `\#12\catcode `\^12\catcode `\_12\catcode `\%12\relax}%
\providecommand \@@startlink[1]{}%
\providecommand \@@endlink[0]{}%
\providecommand \url  [0]{\begingroup\@sanitize@url \@url }%
\providecommand \@url [1]{\endgroup\@href {#1}{\urlprefix }}%
\providecommand \urlprefix  [0]{URL }%
\providecommand \Eprint [0]{\href }%
\providecommand \doibase [0]{https://doi.org/}%
\providecommand \selectlanguage [0]{\@gobble}%
\providecommand \bibinfo  [0]{\@secondoftwo}%
\providecommand \bibfield  [0]{\@secondoftwo}%
\providecommand \translation [1]{[#1]}%
\providecommand \BibitemOpen [0]{}%
\providecommand \bibitemStop [0]{}%
\providecommand \bibitemNoStop [0]{.\EOS\space}%
\providecommand \EOS [0]{\spacefactor3000\relax}%
\providecommand \BibitemShut  [1]{\csname bibitem#1\endcsname}%
\let\auto@bib@innerbib\@empty
\bibitem [{\citenamefont {Baek}\ \emph
  {et~al.}(2026{\natexlab{a}})\citenamefont {Baek}, \citenamefont {Park},
  \citenamefont {Hiyama}, \citenamefont {Noh}, \citenamefont {Yun},
  \citenamefont {Han},\ and\ \citenamefont {Lee}}]{Baek2026ThreeBody}%
  \BibitemOpen
  \bibfield  {author} {\bibinfo {author} {\bibfnamefont {J.}~\bibnamefont
  {Baek}}, \bibinfo {author} {\bibfnamefont {A.}~\bibnamefont {Park}}, \bibinfo
  {author} {\bibfnamefont {E.}~\bibnamefont {Hiyama}}, \bibinfo {author}
  {\bibfnamefont {S.}~\bibnamefont {Noh}}, \bibinfo {author} {\bibfnamefont
  {H.}~\bibnamefont {Yun}}, \bibinfo {author} {\bibfnamefont {K.~C.}\
  \bibnamefont {Han}},\ and\ \bibinfo {author} {\bibfnamefont {S.~H.}\
  \bibnamefont {Lee}},\ }\href {https://arxiv.org/abs/2608.18949} {\bibinfo
  {title} {Three-body forces in the quark model}} (\bibinfo {year}
  {2026}{\natexlab{a}}),\ \bibinfo {note} {revised manuscript under review for
  Prog. Part. Nucl. Phys.},\ \Eprint {https://arxiv.org/abs/2608.18949}
  {arXiv:2608.18949} \BibitemShut {NoStop}%
\bibitem [{\citenamefont {Baek}\ \emph
  {et~al.}(2026{\natexlab{b}})\citenamefont {Baek}, \citenamefont {Park},
  \citenamefont {Hiyama}, \citenamefont {Noh}, \citenamefont {Yun},
  \citenamefont {Han},\ and\ \citenamefont {Lee}}]{Baek2026FlavorDependency}%
  \BibitemOpen
  \bibfield  {author} {\bibinfo {author} {\bibfnamefont {J.}~\bibnamefont
  {Baek}}, \bibinfo {author} {\bibfnamefont {A.}~\bibnamefont {Park}}, \bibinfo
  {author} {\bibfnamefont {E.}~\bibnamefont {Hiyama}}, \bibinfo {author}
  {\bibfnamefont {S.}~\bibnamefont {Noh}}, \bibinfo {author} {\bibfnamefont
  {H.}~\bibnamefont {Yun}}, \bibinfo {author} {\bibfnamefont {K.~C.}\
  \bibnamefont {Han}},\ and\ \bibinfo {author} {\bibfnamefont {S.~H.}\
  \bibnamefont {Lee}},\ }\bibfield  {title} {\bibinfo {title} {Resolving the
  flavor-dependency puzzle in baryon spectroscopy with short-range quark
  three-body forces},\ }\href {https://doi.org/10.1016/j.jspc.2026.100524}
  {\bibfield  {journal} {\bibinfo  {journal} {J. Subatomic Part. Cosmol.}\
  }\textbf {\bibinfo {volume} {6}},\ \bibinfo {pages} {100524} (\bibinfo {year}
  {2026}{\natexlab{b}})}\BibitemShut {NoStop}%
\bibitem [{\citenamefont {Brau}\ \emph {et~al.}(2002)\citenamefont {Brau},
  \citenamefont {Semay},\ and\ \citenamefont
  {Silvestre-Brac}}]{BrauSemaySilvestreBrac2002}%
  \BibitemOpen
  \bibfield  {author} {\bibinfo {author} {\bibfnamefont {F.}~\bibnamefont
  {Brau}}, \bibinfo {author} {\bibfnamefont {C.}~\bibnamefont {Semay}},\ and\
  \bibinfo {author} {\bibfnamefont {B.}~\bibnamefont {Silvestre-Brac}},\
  }\bibfield  {title} {\bibinfo {title} {Unified meson-baryon potential},\
  }\href {https://doi.org/10.1103/PhysRevC.66.055202} {\bibfield  {journal}
  {\bibinfo  {journal} {Phys. Rev. C}\ }\textbf {\bibinfo {volume} {66}},\
  \bibinfo {pages} {055202} (\bibinfo {year} {2002})},\ \Eprint
  {https://arxiv.org/abs/hep-ph/0412176} {arXiv:hep-ph/0412176} \BibitemShut
  {NoStop}%
\bibitem [{\citenamefont {Valcarce}\ \emph {et~al.}(2005)\citenamefont
  {Valcarce}, \citenamefont {Garcilazo},\ and\ \citenamefont
  {Vijande}}]{ValcarceGarcilazoVijande2005}%
  \BibitemOpen
  \bibfield  {author} {\bibinfo {author} {\bibfnamefont {A.}~\bibnamefont
  {Valcarce}}, \bibinfo {author} {\bibfnamefont {H.}~\bibnamefont
  {Garcilazo}},\ and\ \bibinfo {author} {\bibfnamefont {J.}~\bibnamefont
  {Vijande}},\ }\bibfield  {title} {\bibinfo {title} {Constituent quark model
  study of light- and strange-baryon spectra},\ }\href
  {https://doi.org/10.1103/PhysRevC.72.025206} {\bibfield  {journal} {\bibinfo
  {journal} {Phys. Rev. C}\ }\textbf {\bibinfo {volume} {72}},\ \bibinfo
  {pages} {025206} (\bibinfo {year} {2005})},\ \Eprint
  {https://arxiv.org/abs/hep-ph/0507297} {arXiv:hep-ph/0507297} \BibitemShut
  {NoStop}%
\bibitem [{\citenamefont {Oka}\ and\ \citenamefont
  {Takeuchi}(1991)}]{OkaTakeuchi1991}%
  \BibitemOpen
  \bibfield  {author} {\bibinfo {author} {\bibfnamefont {M.}~\bibnamefont
  {Oka}}\ and\ \bibinfo {author} {\bibfnamefont {S.}~\bibnamefont {Takeuchi}},\
  }\bibfield  {title} {\bibinfo {title} {Instanton induced interaction and the
  strange dibaryons},\ }\href {https://doi.org/10.1016/0375-9474(91)90267-A}
  {\bibfield  {journal} {\bibinfo  {journal} {Nucl. Phys. A}\ }\textbf
  {\bibinfo {volume} {524}},\ \bibinfo {pages} {649} (\bibinfo {year}
  {1991})}\BibitemShut {NoStop}%
\bibitem [{\citenamefont {Pepin}\ and\ \citenamefont
  {Stancu}(2002)}]{PepinStancu2002}%
  \BibitemOpen
  \bibfield  {author} {\bibinfo {author} {\bibfnamefont {S.}~\bibnamefont
  {Pepin}}\ and\ \bibinfo {author} {\bibfnamefont {F.}~\bibnamefont {Stancu}},\
  }\bibfield  {title} {\bibinfo {title} {Three-body confinement force in hadron
  spectroscopy},\ }\href {https://doi.org/10.1103/PhysRevD.65.054032}
  {\bibfield  {journal} {\bibinfo  {journal} {Phys. Rev. D}\ }\textbf {\bibinfo
  {volume} {65}},\ \bibinfo {pages} {054032} (\bibinfo {year} {2002})},\
  \Eprint {https://arxiv.org/abs/hep-ph/0105232} {arXiv:hep-ph/0105232}
  \BibitemShut {NoStop}%
\bibitem [{\citenamefont {Papp}\ and\ \citenamefont
  {Stancu}(2003)}]{PappStancu2003}%
  \BibitemOpen
  \bibfield  {author} {\bibinfo {author} {\bibfnamefont {Z.}~\bibnamefont
  {Papp}}\ and\ \bibinfo {author} {\bibfnamefont {F.}~\bibnamefont {Stancu}},\
  }\bibfield  {title} {\bibinfo {title} {Three-body confinement force in a
  realistic constituent quark model},\ }\href
  {https://doi.org/10.1016/j.nuclphysa.2003.07.011} {\bibfield  {journal}
  {\bibinfo  {journal} {Nucl. Phys. A}\ }\textbf {\bibinfo {volume} {726}},\
  \bibinfo {pages} {327} (\bibinfo {year} {2003})},\ \Eprint
  {https://arxiv.org/abs/hep-ph/0206175} {arXiv:hep-ph/0206175} \BibitemShut
  {NoStop}%
\bibitem [{\citenamefont {Takahashi}\ \emph {et~al.}(2002)\citenamefont
  {Takahashi}, \citenamefont {Suganuma}, \citenamefont {Nemoto},\ and\
  \citenamefont {Matsufuru}}]{TakahashiEtAl2002}%
  \BibitemOpen
  \bibfield  {author} {\bibinfo {author} {\bibfnamefont {T.~T.}\ \bibnamefont
  {Takahashi}}, \bibinfo {author} {\bibfnamefont {H.}~\bibnamefont {Suganuma}},
  \bibinfo {author} {\bibfnamefont {Y.}~\bibnamefont {Nemoto}},\ and\ \bibinfo
  {author} {\bibfnamefont {H.}~\bibnamefont {Matsufuru}},\ }\bibfield  {title}
  {\bibinfo {title} {Detailed analysis of the three-quark potential in {SU(3)}
  lattice {QCD}},\ }\href {https://doi.org/10.1103/PhysRevD.65.114509}
  {\bibfield  {journal} {\bibinfo  {journal} {Phys. Rev. D}\ }\textbf {\bibinfo
  {volume} {65}},\ \bibinfo {pages} {114509} (\bibinfo {year} {2002})},\
  \Eprint {https://arxiv.org/abs/hep-lat/0204011} {arXiv:hep-lat/0204011}
  \BibitemShut {NoStop}%
\bibitem [{\citenamefont {Brambilla}\ \emph {et~al.}(2010)\citenamefont
  {Brambilla}, \citenamefont {Ghiglieri},\ and\ \citenamefont
  {Vairo}}]{BrambillaGhiglieriVairo2010}%
  \BibitemOpen
  \bibfield  {author} {\bibinfo {author} {\bibfnamefont {N.}~\bibnamefont
  {Brambilla}}, \bibinfo {author} {\bibfnamefont {J.}~\bibnamefont
  {Ghiglieri}},\ and\ \bibinfo {author} {\bibfnamefont {A.}~\bibnamefont
  {Vairo}},\ }\bibfield  {title} {\bibinfo {title} {Three-quark static
  potential in perturbation theory},\ }\href
  {https://doi.org/10.1103/PhysRevD.81.054031} {\bibfield  {journal} {\bibinfo
  {journal} {Phys. Rev. D}\ }\textbf {\bibinfo {volume} {81}},\ \bibinfo
  {pages} {054031} (\bibinfo {year} {2010})},\ \bibinfo {note} {erratum: Phys.
  Rev. D 107, 019904 (2023)},\ \Eprint {https://arxiv.org/abs/0911.3541}
  {arXiv:0911.3541 [hep-ph]} \BibitemShut {NoStop}%
\bibitem [{\citenamefont {Ekstein}(1960)}]{Ekstein1960}%
  \BibitemOpen
  \bibfield  {author} {\bibinfo {author} {\bibfnamefont {H.}~\bibnamefont
  {Ekstein}},\ }\bibfield  {title} {\bibinfo {title} {Equivalent hamiltonians
  in scattering theory},\ }\href {https://doi.org/10.1103/PhysRev.117.1590}
  {\bibfield  {journal} {\bibinfo  {journal} {Phys. Rev.}\ }\textbf {\bibinfo
  {volume} {117}},\ \bibinfo {pages} {1590} (\bibinfo {year}
  {1960})}\BibitemShut {NoStop}%
\bibitem [{\citenamefont {Coester}\ and\ \citenamefont
  {Polyzou}(1982)}]{CoesterPolyzou1982}%
  \BibitemOpen
  \bibfield  {author} {\bibinfo {author} {\bibfnamefont {F.}~\bibnamefont
  {Coester}}\ and\ \bibinfo {author} {\bibfnamefont {W.~N.}\ \bibnamefont
  {Polyzou}},\ }\bibfield  {title} {\bibinfo {title} {Relativistic quantum
  mechanics of particles with direct interactions},\ }\href
  {https://doi.org/10.1103/PhysRevD.26.1348} {\bibfield  {journal} {\bibinfo
  {journal} {Phys. Rev. D}\ }\textbf {\bibinfo {volume} {26}},\ \bibinfo
  {pages} {1348} (\bibinfo {year} {1982})}\BibitemShut {NoStop}%
\bibitem [{\citenamefont {Polyzou}\ and\ \citenamefont
  {Gl{\"o}ckle}(1990)}]{PolyzouGlockle1990}%
  \BibitemOpen
  \bibfield  {author} {\bibinfo {author} {\bibfnamefont {W.~N.}\ \bibnamefont
  {Polyzou}}\ and\ \bibinfo {author} {\bibfnamefont {W.}~\bibnamefont
  {Gl{\"o}ckle}},\ }\bibfield  {title} {\bibinfo {title} {Three-body
  interactions and on-shell equivalent two-body interactions},\ }\href
  {https://doi.org/10.1007/BF01091701} {\bibfield  {journal} {\bibinfo
  {journal} {Few-Body Syst.}\ }\textbf {\bibinfo {volume} {9}},\ \bibinfo
  {pages} {97} (\bibinfo {year} {1990})}\BibitemShut {NoStop}%
\bibitem [{\citenamefont {Furnstahl}\ \emph {et~al.}(2001)\citenamefont
  {Furnstahl}, \citenamefont {Hammer},\ and\ \citenamefont
  {Tirfessa}}]{FurnstahlHammerTirfessa2001}%
  \BibitemOpen
  \bibfield  {author} {\bibinfo {author} {\bibfnamefont {R.~J.}\ \bibnamefont
  {Furnstahl}}, \bibinfo {author} {\bibfnamefont {H.-W.}\ \bibnamefont
  {Hammer}},\ and\ \bibinfo {author} {\bibfnamefont {N.}~\bibnamefont
  {Tirfessa}},\ }\bibfield  {title} {\bibinfo {title} {Field redefinitions at
  finite density},\ }\href {https://doi.org/10.1016/S0375-9474(00)00687-4}
  {\bibfield  {journal} {\bibinfo  {journal} {Nucl. Phys. A}\ }\textbf
  {\bibinfo {volume} {689}},\ \bibinfo {pages} {846} (\bibinfo {year}
  {2001})},\ \Eprint {https://arxiv.org/abs/nucl-th/0010078}
  {arXiv:nucl-th/0010078} \BibitemShut {NoStop}%
\bibitem [{\citenamefont {Polyzou}(2010)}]{Polyzou2010}%
  \BibitemOpen
  \bibfield  {author} {\bibinfo {author} {\bibfnamefont {W.~N.}\ \bibnamefont
  {Polyzou}},\ }\bibfield  {title} {\bibinfo {title} {Equivalent
  hamiltonians},\ }\href {https://doi.org/10.1103/PhysRevC.82.014002}
  {\bibfield  {journal} {\bibinfo  {journal} {Phys. Rev. C}\ }\textbf {\bibinfo
  {volume} {82}},\ \bibinfo {pages} {014002} (\bibinfo {year} {2010})},\
  \Eprint {https://arxiv.org/abs/1001.4523} {arXiv:1001.4523 [math-ph]}
  \BibitemShut {NoStop}%
\bibitem [{\citenamefont {Shirokov}\ \emph {et~al.}(2007)\citenamefont
  {Shirokov}, \citenamefont {Vary}, \citenamefont {Mazur},\ and\ \citenamefont
  {Weber}}]{Shirokov2007JISP16}%
  \BibitemOpen
  \bibfield  {author} {\bibinfo {author} {\bibfnamefont {A.~M.}\ \bibnamefont
  {Shirokov}}, \bibinfo {author} {\bibfnamefont {J.~P.}\ \bibnamefont {Vary}},
  \bibinfo {author} {\bibfnamefont {A.~I.}\ \bibnamefont {Mazur}},\ and\
  \bibinfo {author} {\bibfnamefont {T.~A.}\ \bibnamefont {Weber}},\ }\bibfield
  {title} {\bibinfo {title} {Realistic nuclear hamiltonian: Ab exitu
  approach},\ }\href {https://doi.org/10.1016/j.physletb.2006.10.066}
  {\bibfield  {journal} {\bibinfo  {journal} {Phys. Lett. B}\ }\textbf
  {\bibinfo {volume} {644}},\ \bibinfo {pages} {33} (\bibinfo {year} {2007})},\
  \Eprint {https://arxiv.org/abs/nucl-th/0512105} {arXiv:nucl-th/0512105}
  \BibitemShut {NoStop}%
\bibitem [{\citenamefont {Shirokov}\ \emph {et~al.}(2012)\citenamefont
  {Shirokov}, \citenamefont {Kulikov}, \citenamefont {Mazur}, \citenamefont
  {Vary},\ and\ \citenamefont {Maris}}]{Shirokov2012DETPET}%
  \BibitemOpen
  \bibfield  {author} {\bibinfo {author} {\bibfnamefont {A.~M.}\ \bibnamefont
  {Shirokov}}, \bibinfo {author} {\bibfnamefont {V.~A.}\ \bibnamefont
  {Kulikov}}, \bibinfo {author} {\bibfnamefont {A.~I.}\ \bibnamefont {Mazur}},
  \bibinfo {author} {\bibfnamefont {J.~P.}\ \bibnamefont {Vary}},\ and\
  \bibinfo {author} {\bibfnamefont {P.}~\bibnamefont {Maris}},\ }\bibfield
  {title} {\bibinfo {title} {Deuteron-equivalent and phase-equivalent
  interactions within light nuclei},\ }\href
  {https://doi.org/10.1103/PhysRevC.85.034004} {\bibfield  {journal} {\bibinfo
  {journal} {Phys. Rev. C}\ }\textbf {\bibinfo {volume} {85}},\ \bibinfo
  {pages} {034004} (\bibinfo {year} {2012})}\BibitemShut {NoStop}%
\bibitem [{\citenamefont {Bogner}\ \emph {et~al.}(2008)\citenamefont {Bogner},
  \citenamefont {Furnstahl},\ and\ \citenamefont
  {Perry}}]{BognerFurnstahlPerry2008}%
  \BibitemOpen
  \bibfield  {author} {\bibinfo {author} {\bibfnamefont {S.~K.}\ \bibnamefont
  {Bogner}}, \bibinfo {author} {\bibfnamefont {R.~J.}\ \bibnamefont
  {Furnstahl}},\ and\ \bibinfo {author} {\bibfnamefont {R.~J.}\ \bibnamefont
  {Perry}},\ }\bibfield  {title} {\bibinfo {title} {Three-body forces produced
  by a similarity renormalization group transformation in a simple model},\
  }\href {https://doi.org/10.1016/j.aop.2007.09.001} {\bibfield  {journal}
  {\bibinfo  {journal} {Ann. Phys.}\ }\textbf {\bibinfo {volume} {323}},\
  \bibinfo {pages} {1478} (\bibinfo {year} {2008})},\ \Eprint
  {https://arxiv.org/abs/0708.1602} {arXiv:0708.1602 [nucl-th]} \BibitemShut
  {NoStop}%
\bibitem [{\citenamefont {Jurgenson}\ \emph {et~al.}(2009)\citenamefont
  {Jurgenson}, \citenamefont {Navr{\'a}til},\ and\ \citenamefont
  {Furnstahl}}]{JurgensonNavratilFurnstahl2009}%
  \BibitemOpen
  \bibfield  {author} {\bibinfo {author} {\bibfnamefont {E.~D.}\ \bibnamefont
  {Jurgenson}}, \bibinfo {author} {\bibfnamefont {P.}~\bibnamefont
  {Navr{\'a}til}},\ and\ \bibinfo {author} {\bibfnamefont {R.~J.}\ \bibnamefont
  {Furnstahl}},\ }\bibfield  {title} {\bibinfo {title} {Evolution of nuclear
  many-body forces with the similarity renormalization group},\ }\href
  {https://doi.org/10.1103/PhysRevLett.103.082501} {\bibfield  {journal}
  {\bibinfo  {journal} {Phys. Rev. Lett.}\ }\textbf {\bibinfo {volume} {103}},\
  \bibinfo {pages} {082501} (\bibinfo {year} {2009})},\ \Eprint
  {https://arxiv.org/abs/0905.1873} {arXiv:0905.1873 [nucl-th]} \BibitemShut
  {NoStop}%
\bibitem [{\citenamefont {Sengbusch}\ and\ \citenamefont
  {Polyzou}(2004)}]{SengbuschPolyzou2004}%
  \BibitemOpen
  \bibfield  {author} {\bibinfo {author} {\bibfnamefont {E.}~\bibnamefont
  {Sengbusch}}\ and\ \bibinfo {author} {\bibfnamefont {W.~N.}\ \bibnamefont
  {Polyzou}},\ }\bibfield  {title} {\bibinfo {title} {Pointlike constituent
  quarks and scattering equivalences},\ }\href
  {https://doi.org/10.1103/PhysRevC.70.058201} {\bibfield  {journal} {\bibinfo
  {journal} {Phys. Rev. C}\ }\textbf {\bibinfo {volume} {70}},\ \bibinfo
  {pages} {058201} (\bibinfo {year} {2004})},\ \Eprint
  {https://arxiv.org/abs/nucl-th/0406042} {arXiv:nucl-th/0406042} \BibitemShut
  {NoStop}%
\bibitem [{\citenamefont {Kim}\ \emph {et~al.}(2026)\citenamefont {Kim},
  \citenamefont {Nam},\ and\ \citenamefont {Baek}}]{KimNamBaek2026Hybrid}%
  \BibitemOpen
  \bibfield  {author} {\bibinfo {author} {\bibfnamefont {S.}~\bibnamefont
  {Kim}}, \bibinfo {author} {\bibfnamefont {G.}~\bibnamefont {Nam}},\ and\
  \bibinfo {author} {\bibfnamefont {J.}~\bibnamefont {Baek}},\ }\href
  {https://arxiv.org/abs/2610.12525} {\bibinfo {title} {Mass and decay angles
  in the {$\Xi_c$--$\Xi_c^{\prime}$} mixing tension: Virtual hybrids and a
  single-direction criterion}} (\bibinfo {year} {2026}),\ \bibinfo {note}
  {revised manuscript under review for Phys. Rev. D},\ \Eprint
  {https://arxiv.org/abs/2610.12525} {arXiv:2610.12525 [hep-ph]} \BibitemShut
  {NoStop}%
\bibitem [{\citenamefont {Hiyama}\ \emph {et~al.}(2003)\citenamefont {Hiyama},
  \citenamefont {Kino},\ and\ \citenamefont
  {Kamimura}}]{HiyamaKinoKamimura2003}%
  \BibitemOpen
  \bibfield  {author} {\bibinfo {author} {\bibfnamefont {E.}~\bibnamefont
  {Hiyama}}, \bibinfo {author} {\bibfnamefont {Y.}~\bibnamefont {Kino}},\ and\
  \bibinfo {author} {\bibfnamefont {M.}~\bibnamefont {Kamimura}},\ }\bibfield
  {title} {\bibinfo {title} {Gaussian expansion method for few-body systems},\
  }\href {https://doi.org/10.1016/S0146-6410(03)90015-9} {\bibfield  {journal}
  {\bibinfo  {journal} {Prog. Part. Nucl. Phys.}\ }\textbf {\bibinfo {volume}
  {51}},\ \bibinfo {pages} {223} (\bibinfo {year} {2003})}\BibitemShut
  {NoStop}%
\bibitem [{\citenamefont {Feshbach}(1958)}]{Feshbach1958}%
  \BibitemOpen
  \bibfield  {author} {\bibinfo {author} {\bibfnamefont {H.}~\bibnamefont
  {Feshbach}},\ }\bibfield  {title} {\bibinfo {title} {Unified theory of
  nuclear reactions},\ }\href {https://doi.org/10.1016/0003-4916(58)90007-1}
  {\bibfield  {journal} {\bibinfo  {journal} {Ann. Phys.}\ }\textbf {\bibinfo
  {volume} {5}},\ \bibinfo {pages} {357} (\bibinfo {year} {1958})}\BibitemShut
  {NoStop}%
\bibitem [{\citenamefont {Bravyi}\ \emph {et~al.}(2011)\citenamefont {Bravyi},
  \citenamefont {DiVincenzo},\ and\ \citenamefont
  {Loss}}]{BravyiDiVincenzoLoss2011}%
  \BibitemOpen
  \bibfield  {author} {\bibinfo {author} {\bibfnamefont {S.}~\bibnamefont
  {Bravyi}}, \bibinfo {author} {\bibfnamefont {D.~P.}\ \bibnamefont
  {DiVincenzo}},\ and\ \bibinfo {author} {\bibfnamefont {D.}~\bibnamefont
  {Loss}},\ }\bibfield  {title} {\bibinfo {title} {Schrieffer--wolff
  transformation for quantum many-body systems},\ }\href
  {https://doi.org/10.1016/j.aop.2011.06.004} {\bibfield  {journal} {\bibinfo
  {journal} {Ann. Phys.}\ }\textbf {\bibinfo {volume} {326}},\ \bibinfo {pages}
  {2793} (\bibinfo {year} {2011})},\ \Eprint {https://arxiv.org/abs/1105.0675}
  {arXiv:1105.0675 [quant-ph]} \BibitemShut {NoStop}%
\end{thebibliography}%
\end{document}